\documentclass[12pt]{article}
\usepackage{amsmath,amssymb,,mathrsfs,a4wide,comment}
\usepackage{jheppub}
\usepackage{bbm}

\newcommand{\jtheta}[2]{\genfrac{[}{]}{0pt}{}{#1}{#2}}

\newcommand{\dr}{\mathrm{d}}

\newcommand{\unit}{\mathbbm{1}}

\begin{document}

\preprint{Imperial-TP-2019-CH-04}

\title{Heterotic/type II Duality and Non-Geometric Compactifications}
\author[a,b]{Y. Gautier}
\author[c]{C.M. Hull,}
\author[a,b]{D. Isra\"el}

\affiliation[a]{LPTHE, UMR 7589, Sorbonne Universit\'es, UPMC Univ. Paris 06, 4 place Jussieu, Paris, France}
\affiliation[b]{CNRS, UMR 7589, LPTHE, F-75005, Paris, France}
\affiliation[c]{The Blackett Laboratory, Imperial College London, Prince Consort Road, London SW7 2AZ, United Kingdom}

\emailAdd{gautier@lpthe.jussieu.fr}
\emailAdd{c.hull@imperial.ac.uk}
\emailAdd{israel@lpthe.jussieu.fr}

\null\vskip10pt
\abstract{ We present a new class of dualities relating non-geometric Calabi-Yau 
compactifications of  type II string theory to T-fold compactifications of the heterotic string, both preserving four-dimensional 
$\mathcal{N}=2$ supersymmetry. The non-geometric Calabi-Yau space is a $K3$ fibration over $T^2$ with non-geometric monodromies 
in the duality group $O(\Gamma_{4,20})$; this is dual to a heterotic reduction on a $T^4$ fibration over $T^2$ with  the $O(\Gamma_{4,20})$ monodromies
  now  viewed as heterotic T-dualities. At a point in moduli space which is a minimum of the scalar potential, 
the type II compactification becomes an asymmetric Gepner model and the monodromies become automorphisms involving mirror symmetries, while the heterotic dual is an asymmetric toroidal orbifold. We generalise previous constructions to ones in which the automorphisms  are not of prime order.
The type II construction is perturbatively consistent, but the naive heterotic dual is not modular invariant. Modular invariance on the 
heterotic side is achieved by  including  twists in the circles dual to the winding numbers round the $T^2$, and this in turn 
introduces non-perturbative phases depending on  NS5-brane charge in the type II construction.
}

                              
\maketitle

\section{Introduction}

Perturbative type IIA vacua preserving $\mathcal{N}=2$ supersymmetry in four dimensions can be obtained from compactifications on 
Calabi-Yau three-folds (CY$_3$), or from orbifold or Gepner-point limits of these. In these cases the underlying worldsheet $(c,\bar c)= (9,9)$ conformal field theory (CFT) has an 
extended $(2,2)$  superconformal symmetry and  both left- and right-moving R-charges are integer-valued. From the worldsheet perspective, four of the eight space-time 
supercharges come from the left-movers and the other four from the right-movers.

There exists, however, another  possibility, in which {\it all eight}   of the supercharges come from, say, the left-movers. 
This happens in a $ (2,2)$ superconformal field theory (SCFT) with 
integer-valued R-charges from the left-movers but no integer-valued R-charges from the right-movers;  examples 
of this arise in   free-fermion constructions or asymmetric toroidal orbifolds (see {\it e.g.}~\cite{Sen:1995ff}).  A large class 
of   non-geometric models  with all supercharges arising from left-movers  based on Calabi-Yau compactifications in the 
Landau-Ginsburg regime were recently studied in~\cite{Israel:2013wwa,Israel:2015efa,Blumenhagen:2016axv}, related 
to older works~\cite{Intriligator:1990ua,Schellekens:1989wx}. These models  have a   volume modulus for the target space which is fixed by the construction, so that one 
cannot continuously take a   large-volume limit, 
and are intrinsically non-geometric with the number 
of massless moduli  typically being very small.

In~\cite{Hull:2017llx}, 
a new class of ``compactifications'' of type II strings 
to four dimensions was found, based on the work on Landau-Ginzburg models 
mentioned above, in which all  eight supersymmetries   come from the left-moving sector. 
The starting point is type IIA string theory compactified on K3 with duality symmetry $O(\Gamma_{4,20})$, which is the group of isometries of the 
charge lattice $\Gamma_{4,20}$. This is then followed by a duality-twisted compactification on $T^2$ with an $O(\Gamma_{4,20})$ monodromy round each circle. 
The conditions on the monodromies for this to give a stable four-dimensional Minkowski vacuum preserving  $\mathcal{N}=2$ supersymmetry were found and 
an explicit algebraic geometric construction of such monodromies was given. These give non-geometric backgrounds of type II string theory
giving four-dimensional Minkowski vacua preserving the same amount of supersymmetry as
    Calabi-Yau compactifications  and, for this reason, such backgrounds  were referred to as     non-geometric 
Calabi-Yau spaces. As the monodromies involve mirror transformations, the non-geometric internal spaces are 
mirror-folds~\cite{Hull:2004in}. 

The two monodromies $\gamma_1,\gamma_2\in O(\Gamma_{4,20})$ satisfying the conditions for $\mathcal{N}=2$ Minkowski vacua 
are necessarily of finite order, $p_1, p_2$, with $\gamma_1^{p_1} =\gamma_2^{p_2}=1$ for some integers $p_1,p_2$. In~\cite{Hull:2017llx}, 
constructions were given for prime orders $p_1,p_2$ but recent mathematical  results~\cite{mirror,mirror2} allow us to extend the constructions 
of~\cite{Hull:2017llx} to non-prime integers $p_1,p_2$. Each duality $\gamma$ satisfying the conditions above has a 
fixed locus (i.e. locus of fixed points) in the K3 moduli space~\cite{Hull:2017llx}, so that $\gamma$ is an automorphism of the 
 K3 CFT at any point in moduli space that is on the fixed locus. Moreover, the K3 surface $X$ admitting the automorphism $\gamma$ is an algebraic surface, with a mirror 
algebraic K3 surface $\tilde X$, such that the action of $\gamma$ on $X$ can be understood as the composition of four transformations: 
a diffeomorphism of $X$ followed by the mirror map to $\tilde X$, then a diffeomorphism of $\tilde X$ followed 
by the inverse mirror map back to $X$. In the K3 CFT at a Landau-Ginzburg orbifold point, the diffeomorphism of $\tilde X$ appears as a discrete torsion. 
In~\cite{Hull:2017llx}, such automorphisms were referred to as  {\it mirrorred automorphisms}; it is 
striking that they involve transformations of both the K3 surface and its mirror.

For the twisted reduction, $\gamma_1,\gamma_2$ must commute and, if there is to be a Minkowski vacuum, the intersection of their fixed loci must be non-empty.  The orbifold is by transformations $(\gamma_1,t _1)$, 
$(\gamma_2,t _2)$ where the automorphism $\gamma_i$ of degree $p_i$ is combined with a shift $t _i$ on the $i$'th circle of the $T^2$ by $2\pi / p_i$ ($i=1,2$). 
Then at a fixed point the twisted reduction reduces to a freely-acting asymmetric orbifold of the $K3\times T^2$ compactification, resulting in the simplest cases in 
the asymmetric Gepner models of~\cite{Israel:2013wwa}.

In this work we will focus on the heterotic duals of these constructions, using the  duality between type IIA string theory 
compactified on K3 and heterotic string theory compactified on $T^4$~\cite{Hull:1994ys}. Starting from the $\mathcal{N}=4$ duality in 
four dimensions between the type IIA string on $K3\times T^2$ and 
the heterotic string  on $T^6$, one can reach $\mathcal{N}=2$ dual pairs through (freely-acting)  orbifolds preserving half of 
the supersymmetry. An important example, the construction of Ferrara, Harvey, Strominger and Vafa (FHSV)~\cite{Ferrara:1995yx}, relates the type  IIA  string compactified on the 
Enriques Calabi-Yau three-fold to an asymmetric toroidal orbifold of the heterotic string. More generally, it is expected that type IIA compactified on a $K3$-fibred CY$_3$ 
with a compatible elliptic fibration is non-perturbatively dual to a heterotic string compactification on $K3 \times T^2$; see~\cite{Aspinwall:1996mn} 
for a review. Our models extend this to non-geometric dual constructions.

In the six-dimensional heterotic/type IIA duality, the $O(\Gamma_{4,20})$ duality symmetry group of the type IIA string compactified on $K3$ 
is identified as the $O(\Gamma_{4,20})$ T-duality symmetry group of the heterotic 
string, for which $\Gamma_{4,20}$ is the Narain lattice. Then the duality-twisted reduction on $T^2$ with monodromies $\gamma_1,\gamma_2\in O(\Gamma_{4,20})$ has a heterotic realisation as 
a T-fold~\cite{Hull:2004in,Dabholkar:2002sy,Dabholkar:2005ve} with T-duality monodromies  -- it is a ``compactification'' of 
the heterotic string on a non-geometric space that has a  
fibration of $T^4$ CFTs over a $T^2$ base, with T-duality transition functions.
Then the heterotic/type IIA duality maps the non-geometric Calabi-Yau mirror-fold reduction of type IIA to a T-fold reduction of the heterotic string.
At a fixed point in moduli space
(a point preserved by  both $\gamma_1,\gamma_2$), the heterotic T-fold reduces to an asymmetric orbifold of the heterotic string on $T^6$ by the transformations $(\gamma_1,t _1)$, 
$(\gamma_2,t _2)$ consisting of $O(\Gamma_{4,20})$ T-duality transformations on $T^4$ combined with shifts on $T^2$. The K3 CFT at the fixed point in moduli space gives no enhanced gauge 
symmetry, so the corresponding $T^4$  heterotic compactification also has no enhanced gauge symmetry --~instead it has enhanced discrete symmetry as in~\cite{Harvey:2017xdt}.

In a recent article~\cite{Harvey:2017xdt}, Harvey and Moore made the following point: {\it ``It is not, a priori, obvious that heterotic/type II duality should apply 
to asymmetric orbifolds of  the heterotic string''}.  Indeed, while the FHSV model provides  an example of such a  dual to an  asymmetric heterotic orbifold, 
no general statement appears to have been made so  far. It is usually assumed that  the type IIA side of such a  duality should 
involve a Calabi-Yau three-fold.  Here we   show that in many cases  an asymmetric  orbifold of the heterotic string has a type 
IIA dual that is a non-geometric compactification, an orbifold of the type IIA string on $K3\times T^2$  by non-geometric symmetries.

Remarkably, while the IIA construction is a consistent construction  for the perturbative type IIA string,
the naive heterotic dual  is not perturbatively consistent -- it is not modular invariant.
 The perturbative heterotic construction can be modified to obtain modular invariance, 
but via duality this then introduces non-perturbative modifications to the type IIA construction.
This complies with the adiabatic argument put forward in~\cite{Vafa:1995gm} to relate non-perturbative 
dualities with different amounts of supersymmetry. Such non-perturbative modifications were also seen in the 
FHSV model.
As we shall see, for the asymmetric orbifolds discussed here, modular invariance of the heterotic models 
is only obtained if the shifts  on the two-torus 
are combined with shifts  on the T-dual torus,
corresponding to introducing phases dependent on the string winding numbers on the two-torus.
Under heterotic/type II duality, the fundamental heterotic string is mapped to a IIA NS5-brane wrapping K3, so that new heterotic 
phases are mapped to phases dependent on NS5 wrapping numbers in the IIA string. These NS5-brane contributions give non-perturbative 
modifications to the non-geometric Calabi-Yau construction. The corresponding non-perturbative corrections to the prepotential governing 
the vector moduli space geometry in the low-energy type IIA effective action will be analyzed in a forthcoming publication~\cite{ghi}.

In this paper we will mostly focus on the case when the second twist $\gamma_2$ and shift $t_2$  are trivial. Having one non-trivial twist $\gamma_1$ is sufficient to break 
supersymmetry to $\mathcal{N}=2$ and give an interesting class of models. This can be thought of as a duality-twisted reduction on a circle  to five 
dimensions with monodromy $\gamma_1$ followed by a conventional (untwisted) compactification on a further circle. This is sufficient for most of 
our purposes; the generalisation to two twists is straightforward and will be discussed briefly.

Once the heterotic dual has been found, non-perturbative aspects of the theory can be probed. We will study the perturbative heterotic BPS states
that are 
dual  to  type IIA bound states of NS5-branes (wrapping a one-cycle of the two-torus and the $K3$ fibre) and momentum states on the $T^2$
by computing the generating function for the helicity supertraces.  The map between BPS states is, in a way, easier to understand than in standard cases of 
$\mathcal{N}=2$ heterotic/type II dualities as there are no D-branes bound-states to take into account in the present context.

The plan of this paper is as follows. In section~\ref{sec:typeII} we summarize the general construction of non-geometric Calabi-Yau 
backgrounds and the results from~\cite{Hull:2017llx} that are needed for this paper. In section~\ref{sec:auty} we briefly review 
the mirrored automorphisms introduced in~\cite{Hull:2017llx} and the
way they lead to
 the construction of isometries of the $\Gamma_{4,20}$ lattice that satisfy the conditions 
 needed for a non-geometric Calabi-Yau background;
  these details are not needed for the rest of the paper and can be skipped by the impatient reader. In section~\ref{sec:hetdual} 
we find the heterotic dual of the non-geometric Calabi-Yau type IIA models. 
Section~\ref{sec:bps} discusses some of the BPS states that arise and calculates the corresponding indices.
In section~\ref{sec:mirror} we present the consequences of perturbative heterotic consistency in the type IIA duality frame. Section~\ref{sec:dualmir} is devoted 
to a duality-covariant analysis of our models in four and five dimensions and of the FHSV model, allowing us to construct further dual forms of these models. 
Finally conclusions %and avenues for future work 
are presented in section~\ref{sec:concl}.

\section{Non-Geometric Calabi-Yau Backgrounds}
\label{sec:typeII}

In this section we summarize the construction of non-geometric Calabi-Yau backgrounds from~\cite{Hull:2017llx}. Type IIA 
string theory on K3 has a world-sheet formulation as a supersymmetric non-linear sigma model with K3 target space, with a B-field given by a closed 2-form; this defines a superconformal field theory.
The moduli space of such non-linear sigma-models on $K3$ surfaces is given by~\cite{Seiberg:1988pf,Aspinwall:1994rg}
\begin{equation}
  \label{eq:NLSMmoduli}
\mathcal{M}_\sigma \cong O(\Gamma_{4,20}) \backslash O(4,20) \slash O(4) \times O(20)\, .
\end{equation}
The isometry group of the branes charge lattice $\Gamma_{4,20}$ (which is also the lattice of total cohomology of the K3 surface and is given 
by eq.~(\ref{eq:totalcoh}) below) is denoted by $O(\Gamma_{4,20})$. The string theory duality symmetry group 
must preserve the charge lattice hence is given by $O(\Gamma_{4,20})$,  with a natural action on the moduli space.
It acts on the perturbative theory through automorphisms of the conformal field theory on K3, but this extends to an action on the non-perturbative theory.

Locally, the moduli space can be decomposed as 
\begin{equation}
O(4,20) \slash [O(4) \times O(20)]\cong O(3,19) \slash [ O(3) \times O(19)] \times \mathbb{R}^{22} \times \mathbb{R}^+\, .
\end{equation}
The first term on the right hand side is the moduli space
  of Ricci-flat K\"ahler  metrics on $K3$, 
  the second is  the 
  cohomology group $H^2(K3; \mathbb{R})$    giving moduli space of flat B-fields and the last term is the volume modulus of 
the K3 surface.  The duality group contains a geometric subgroup 
$O(\Gamma_{3,19}) \ltimes  \mathbb{Z}^{22}$ generated by large diffeomorphisms of the surface in $O(\Gamma_{3,19})$ and 
integral shifts of the B-field, i.e. shifts of $B$ by a 2-form representing   an integral cohomology class.
The remaining dualities are non-geometric in character. At certain special points in the moduli space corresponding to 
algebraic K3 surfaces,  
 mirror symmetry provides an extra generator 
of order two  which, 
together with the geometric subgroup
generates the full duality group  $O(\Gamma_{4,20})$~\cite{Nahm:2001kh,Aspinwall:1994rg}. 

There is a continuous action of the group $O(4,20)$ on the moduli space and hence on the type IIA string compactified  on K3.
The type IIA string compactified  on K3 can be further compactified on  $T^2$ with duality twists through an ansatz in which the 
dependence of all fields on the toroidal coordinates $y^1,y^2$ is given by a $y^i$-dependent $O(4,20)$  transformation:
\begin{equation}
\label{SchSch}
g_1(y^1)  = e^{N_{1}  y^1} \ , \quad 
g_2(y^2)  = e^{N_{2}  y^2} \, .
\end{equation}
for two commuting Lie algebra generators $N_1,N_2$.
Then the monodromies are
\begin{equation}
\label{eq:monod}
\gamma_1=g_1(0)^{-1}g_1(2\pi R_1)= e^{2\pi R_1 \, N_{1} }, \qquad
\gamma_2=g_2(0)^{-1}g_2(2\pi R_2)= e^{2\pi R_2 \, N_{2} }
\end{equation}

This compactification has a low energy effective action given by a gauged ${\cal N}=4$ supergravity theory in four dimensions~\cite{ReidEdwards:2008rd,Hull:2017llx}.
The scalar potential will have a global minimum with zero energy, giving a Minkowski space vacuum, if the monodromies are elliptic, i.e.
they are $O(4,20)$-conjugate to elements of the compact subgroup $O(4)\times O (20)$. Then each monodromy $\gamma_i$ is an element of the discrete group $O(\Gamma_{4,20})\subset O(4,20)$ such that there 
are  $U_i\in O(4,20)$ and $M_i \in O(4)\times O(20)$ with 
\begin{equation}
\label{eq:gamM}
\gamma_i =  U_i M_i U^{-1}_i \ ; \qquad U_i\in O(4,20)\ , \qquad M_i\in O(4)\times O (20)\ , \qquad i=1,2. 
\end{equation}
Each monodromy is of finite order, i.e. there are integers $p_1,p_2$ such that 
\begin{equation}
\gamma _1 ^{p_1} =\gamma _2 ^{p_2} =1\, ,
\end{equation}
and each will have a fixed locus in the Teichmuller space $O(4,20) \slash O(4) \times O(20)$. We will denote the coset containing $g\in O(4,20)$ as $(g)$.
As a rotation $O(4)\times O (20)$  has a fixed point at the origin $(1)$, the fixed loci of $\gamma_1,\gamma_2$ will spanned by $(U_1), (U_2)$ with $U_i$ satifying~(\ref{eq:gamM}).

For there to be a global minimum of the potential, the intersection of the two fixed loci should be non-empty, and  we will take $U_1=U_2\equiv U$ inside the intersection. 
Then there will be a  minimum of the potential   be at $(U)$ and each monodromy $\gamma_i$ is an $O(\Gamma_{4,20})$ transformation conjugate to a rotation $M_i\in O(4)\times  O(20)$. Conjugating 
both monodromies by the same element $V$ of $O(\Gamma_{4,20})$ then takes 
\begin{equation}
\gamma_i \mapsto \gamma_i  '= V\gamma_i  V^{-1}
\end{equation}
with a point of the fixed locus now at $(VU)$.
In this way, one can always arrange for an element of the fixed locus to be in any given fundamental domain of the Teichmuller space.

Regarding $g$ as a $24\times 24$ matrix    acting in the fundamental representation of $O(4,20)$, the left coset $O(4,20) \slash O(4) \times O(20)$
can be parameterised by the \lq generalised metric'
 \begin{equation}
{\cal H}(g)= g^t g \, .
\end{equation}
The group $O(4,20)$ acts on this by
 \begin{equation}
{\cal H}\mapsto k^t{\cal H} k, ~ k \in O(4,20)  \, .
\end{equation}
Then the stabiliser of a point $(g_0)\in
O(4,20) \slash O(4) \times O(20)
$
is the subgroup $H_0\subset
O(4,20) $ preserving ${\cal H}_0= {\cal H}(g_0)$
given by
\begin{equation}
H_0=\{g\in O(4,20) : g^t {\cal H}_0g =
{\cal H}_0  \} \,  .
\end{equation}
At the identity, $g_0=\unit $,  ${\cal H}_0=\unit $ and $H_0$ is the standard $ O(4) \times O(20)$ subgroup
\begin{equation}
H(\unit )=\{g\in O(4,20) : g^t  g =\unit \}
\end{equation}
while for general $g_0$, $H_0$ is a conjugate $ O(4) \times O(20)$ subgroup
\begin{equation}
H_0=
 g_0^{-1} H(\unit )g_0= \{ g\in O(4,20) : g= 
  g_0^{-1}  k g_0
 , ~ k\in H(\unit ) \} \, .
\end{equation}
We will write this as
\begin{equation}
H_0= O(4)_0\times O(20)_0 
\end{equation}
where $O(4)_0$ is conjugate to the standard $O(4)$ and $O(20)_0 $ is conjugate to the usual $O(20)$:
\begin{subequations}
\label{grops}
\begin{align}
O(4)_0 &= 
  \{ g\in O(4,20) : g= 
  g_0^{-1}  k g_0
 , ~ k\in O(4) \subset H(\unit ) \},
 \label{grops_a} \\
 O(20)_0 &= 
  \{ g\in O(4,20) : g= 
  g_0^{-1}  k g_0
 , ~ k\in O(20) \subset H(\unit ) \} 
\, .
 \label{grops_b}
\end{align}
\end{subequations}

As a result, any automorphism at $(g_0)$ must be in the  $ O(4) \times O(20)$ subgroup $H_0$, and so the monodromies $\gamma_1,  \gamma_2$ must be in 
\begin{equation}
O(\Gamma_{4,20}) \cap H_0
\end{equation}
and we see that (\ref{eq:gamM}) is satisfied with 
\begin{equation}
U^{-1}  = 
g_0
\end{equation}
for an  $ O(4) \times O(20)$ matrix $M\in H(\unit )$.

The models that we consider should furthermore preserve eight supercharges in four dimensions. Taking the $O(4)$ part of the  rotation $M$  to be in 
 \begin{equation}
SO(4)\sim 
\frac {SU(2)_L\times SU(2)_R} {\mathbb{Z}_2}\, , 
\end{equation}
the condition for the reduction to preserve 8 of the 16 supersymmetries and so to give ${\cal N}=2$ supersymmetry in four dimensions is that the rotation $M$ is in 
$SU(2)_L\times O(20)$ or $SU(2)_R\times O(20)$.

Then the twisted reduction  giving 
an ${\cal N}=2$ supersymmetric Minkowski vacuum in four dimensions
consists of a duality  twist with monodromy $\gamma_1$ of order $p_1$ on the $y^1$ circle and a twist  of $\gamma_2 $ of order $p_2$ on the $y^2$ circle with
\begin{equation}
\gamma _i =  U M _iU^{-1}; \qquad U\in O(4,20), \qquad M_i \in SU(2)\times O(20)
\end{equation}
At some fixed point in moduli space, the reduction becomes an 
 orbifold  by transformations $(\gamma_1,t _1)$, 
$(\gamma_2,t _2)$ where  $t _i$ is a shift on the $i$'th circle of order $p_i$
\begin{equation}
t _i: y^i\to y^i + 2 \pi /p_i
\end{equation}
and the twisted reduction reduces to  a freely-acting asymmetric
 ${\mathbb{Z}}_{p_1} 
 \times
 {\mathbb{Z}}_{p_2}$ 
  orbifold of the $K3\times T^2$ compactification.

 An interesting  class of models is that in which one of the monodromies is trivial, $\gamma_2=1$.
 Then we have a twisted reduction on one circle with monodromy $\gamma_1$ and a standard (untwisted) reduction on the other circle.
 This is sufficient to break the supersymmetry to  ${\cal N}=2$ and gives a simple class of models that captures many of the features we want to study.
 We will focus on the implications of a single twist here; the second twist would be treated similarly and doesn't qualitatively change the physics, 
but leads to a more general mass spectrum, as discussed in~\cite{Hull:2017llx}.
 
 For a single twist $\gamma$ conjugate to a rotation $M\in SO(4)\times SO(20)$ by  (\ref{eq:gamM}), the $SO(4)$ rotation is characterised by two angles $s_1,s_2$ and the 
 $SO(20)$ rotation is characterised by ten angles
 $r_1,\dots r_{10}$. For supersymmetry, the $SO(4)$ angles must satisfy $s_1= \pm s_2$~\cite{Hull:2017llx}.
For any admissible twist, $\gamma$ satisfying our conditions, $V\gamma_i  V^{-1}$ will also be an admissible twist for all $V\in O(\Gamma_{4,20})$.
 Changing from $\gamma$ to $V\gamma_i  V^{-1}$  will move a fixed point in Teichmuller space from $(U)$ to $(VU)$. As the volume of the K3 is one of the moduli, this change of 
representative can change the volume of the K3; all such theories related in this way are physically equivalent as they are related by 
dualities.% and the overall volume of the K3 can be large or small depending on the choice of duality frame (or fundamental domain).

It is a rather non-trivial problem to find  $24\times 24$ integer-valued matrices representing elements of $O(\Gamma_{4,20})$ that are conjugate to $SU(2)\times O(20) $ rotations.
In~\cite{Hull:2017llx}, an explicit construction was given. 
The starting point was finding a special algebraic K3 surface with a geometric automorphism $\sigma$ of order $p$, and then constructing from this an automorphism $\hat \sigma $ 
of the K3 conformal field theory whose action $\hat \sigma ^*$  on the lattice $\Gamma_{4,20}$
satisfied  all the conditions above, and so taking $\gamma=\hat \sigma ^*$ gives the construction of our non-geometric Calabi-Yau space.
These are the mirrored automorphisms and their construction is reviewed in the next section.
Then the same $\gamma$ will then  be used in the dual heterotic construction in section 4, with the $O(\Gamma_{4,20})$ transformation realised as an element of the heterotic T-duality group.

\section{Mirrored Automorphisms}
\label{sec:auty}

In this section we review the construction of mirrored automorphisms
of K3 from~\cite{Hull:2017llx}, in which the K3  is chosen to be an algebraic K3 surface, {\it i.e.} a hypersurface in a weighted projective space.
We start by recalling the description of the non-geometric duality symmetries as lattice isometries.

 The integral second cohomology of a K3 surface is isomorphic to an even self-dual lattice of signature $(3,19)$. Up to isometries it is unique and given by
\begin{equation}
\Gamma_{3,19} \cong E_8 \oplus E_8 \oplus U \oplus U \oplus U\, , 
\end{equation}
where $E_8$ is the negative-definite lattice associated with the $E_8$ Dynkin diagram and $U$ the unique even self-dual 
lattice of signature $(1,1)$. In the string theory context it is natural to consider the lattice of 
total cohomology, of signature $(4,20)$,  using the natural pairing between four-forms and zero-forms:
\begin{equation}
\label{eq:totalcoh}
\Gamma_{4,20} \cong \Gamma_{3,19} \oplus U \, .
\end{equation} 
This is also isomorphic to the D-branes charge lattice of the type IIA string compactified on K3. The isometry group of this lattice is  $O(\Gamma_{4,20})$. 
The moduli space of non-linear sigma-models on K3 surfaces is given by~(\ref{eq:NLSMmoduli}) and the action of a duality on the moduli 
space corresponds to an element of the isometry group of the lattice, $O(\Gamma_{4,20})$. The fixed points of those 
transformations are associated with orbifold CFTs.

\subsection{Non-symplectic K3 Automorphisms}

An order $p$ non-symplectic automorphism $\sigma_p$ of a $K3$ surface $X$  is an automorphism that acts on the holomorphic two-form $\omega (X)$ as 
\begin{equation}
\omega (X) \mapsto \sigma_p^{\, \star}\, \omega (X) = \zeta_p \, \omega (X)\, ,
\end{equation}
where $\zeta_p$ is a primitive $p$-th root of unity\footnote{$\zeta$ is a primitive $p$'th root of unity 
if $\zeta^p=1$ but $\zeta^n\ne 1$ for $n=1,\dots p-1$.}. As such, in the string theory context, an orbifold by 
a non-symplectic $K3$ automorphism breaks all space-time supersymmetry.

The action of $\sigma_p$ on the cohomology classes singles out an invariant sublattice $S^\mathfrak{Q} (\sigma_p)$ of $\Gamma_{4,20}$, 
which is of signature $(2,\rho)$ with $\rho \in \{1,\ldots,20\}$. The automorphism acts as a non-trivial isometry of the orthogonal complement $T(\sigma_p)$ of 
$S^\mathfrak{Q} (\sigma_p)$, $$T(\sigma_p) := S^\mathfrak{Q} (\sigma_p)^\perp \cap \Gamma_{4,20},$$ which is of signature $(2,20-\rho)$. 

An example of a K3 surface admitting an order 3 non-symplectic automorphism that we will use to illustrate the general idea is provided by the hypersurface
\begin{equation}
\label{eq:23824}
w^2 + x^{3} + y^{8} + z^{24} =0 \ \subset \ \mathbb{P}_{[12, 8, 3, 1]}\, .
\end{equation}
The order 3 automorphism is then simply defined by $\sigma_3 \, : \ x \mapsto e^{2i\pi/3}\, x$. 
The invariant sublattice of $\Gamma_{4,20}$ w.r.t. the action of $\sigma_3$ and its orthogonal complement are given in this case by 
\begin{equation}
\label{eq:TSdecomp}
S^{\mathfrak{Q}} (\sigma_3) = E_6 \oplus U  \oplus U \ , \quad T(\sigma_3) = E_8 \oplus A_2 \oplus U \oplus U\, .
\end{equation}
where $E_6$ and $A_2$ are the negative-definite lattices associated with the corresponding Dynkin diagrams. The action of $\sigma_3$ on the vector space 
$T(\sigma_3) \otimes \mathbb{R}$ corresponds to an element of the orthogonal group $O( T(\sigma_3) )$ that can be found explicitly in~\cite{Hull:2017llx}.

\subsection{Mirrored K3 Automorphisms}

For each non-symplectic $K3$ automorphism of the type described above, there is a corresponding {\it mirrored automorphism} that we will   describe below. 
Supersymmetric  non-geometric orbifolds of type IIA on $K3$ can be obtained by 
orbifolding by such mirrored automorphisms, even though the 
orbifold by the corresponding non-symplectic $K3$ automorphism breaks all supersymmetry.
 For simplicity we will restrict the discussion to the hypersurface~(\ref{eq:23824}); the general case is described in~\cite{Hull:2017llx}. 

The mirror of the $K3$ surface~(\ref{eq:23824}), using the Greene-Plesser map~\cite{Greene:1990ud}, is given by an orbifold of a similar hypersurface
\begin{equation}
\label{eq:23824m}
\tilde w^2 + \tilde x^{3} + \tilde y^{8} + \tilde z^{24} =0 \ \subset \ \mathbb{P}_{[12, 8, 3, 1]}
\end{equation}
by a discrete symmetry group $G\cong \mathbb{Z}_2$ 
generated by 
\begin{equation}
g \, : \left\{ \begin{array}{ccc} \tilde w & \mapsto & - \tilde w \\ \tilde y & \mapsto   & - \tilde y \end{array}\right.  
\end{equation}
This mirror surface also admits an order three automorphism $\tilde{\sigma}_3$, which acts in a similar same way to 
$\sigma_3$, with $\tilde{\sigma}_3 \, : \ \tilde x \mapsto e^{2i\pi/3} \, \tilde x$. However, 
the invariant sublattice for $\tilde{\sigma}_3$ and its orthogonal complement in $\Gamma_{4,20}$ are
now
\begin{equation}
S^{\mathfrak{Q}} (\tilde{\sigma}_3) = E_8 \oplus A_2 \oplus U \oplus U \ , \quad T(\tilde{\sigma}_3) =   E_6 \oplus U  \oplus U\, .
\end{equation}
Comparing with the corresponding sublattices~(\ref{eq:TSdecomp}) for the original  surface~(\ref{eq:23824}), we see that the two sublattices have been interchanged.

A crucial statement,  proved in~\cite{Hull:2017llx}, is that the automorphism $\sigma_3$ of the surface~(\ref{eq:23824})
and the corresponding automorphism $\tilde{\sigma}_3$ of the mirror K3 surface act on orthogonal vector sub-spaces of $\Gamma_{4,20} \otimes \mathbb{R}$:  
$\sigma_3$ acts non-trivially on
$T(\sigma_3) \otimes \mathbb{R}$ and 
$\tilde{\sigma}_3$
acts non-trivially on
$T(\tilde{\sigma}_3) \otimes \mathbb{R}$. The diagonal action of the  corresponding elements 
of $O(T(\sigma_3))$ and $O(T(\tilde{\sigma}_3))$ then lifts to an isometry of the full lattice $\Gamma_{4,20}$, and so provides an element of $O(\Gamma_{4,20})$ of order three. 

In general, the action of a non-symplectic automorphism $\sigma_p$ of order $p$ gives an isometry in $O(T(\sigma_p))$ and the action of $\tilde{\sigma}_p$ on the 
mirror surface gives an isometry in $O(T(\tilde{\sigma}_p))$; their diagonal action is  then lifted to an isometry of $\Gamma_{4,20}$ of order $p$. One 
associates to this isometry the action of a mirrored automorphism $\hat{\sigma}_p$, which can be thought of as 
\begin{equation}
\label{eq:mirrored_aut_dec}
\hat{\sigma}_p = \mu ^{-1} \circ \tilde{\sigma}_p \circ \mu \circ \sigma_p \, ,  
\end{equation}  where  $\mu$ is the mirror map from the original K3 to its mirror.  

In~\cite{Hull:2017llx} a proof of this statement was given for all non-symplectic automorphisms 
of prime order $p\in \{2,3,5,7,13\}$ using theorems proven in~\cite{Artebani2014758,comparin2014} but, crucially, recent mathematical results 
extend this picture to all allowed orders,  including those that are not prime numbers~\cite{mirror,mirror2}.

As was discussed in~\cite{Israel:2015efa,Hull:2017llx}, mirrored automorphisms preserve all space-time supercharges coming 
from the left-movers on the worldsheet, while  projecting out all that come from the right-movers; this would not be possible 
with geometric automorphisms.

\subsection{Non-Geometric \texorpdfstring{$K3 \times T^2$}{K3 x T2} Orbifolds}

The fixed points of  mirrored automorphisms, {\it i.e.} the K3 CFTs that are invariant under both the automorphism $\sigma_p$ and the automorphism $\tilde{\sigma}_p$ of 
a mirror pair, can be orbifolded by the automorphism. Of particular interest are certain models for which there is a duality frame in which the K3 surface has small volume in string units. These give
Landau-Ginzburg (LG) orbifolds~\cite{Intriligator:1990ua} which are special points in the moduli space  of non-linear models on algebraic K3 surfaces at small volume; 
when the polynomial defining the surface is of the Fermat type, as in~(\ref{eq:23824}), they can be described as Gepner models~\cite{Gepner:1987qi} and are explicitly solvable CFTs. 

In this framework, the cyclic group generated by the automorphism $\widetilde{\sigma}_p$ of the mirror surface is an order $p$ subgroup of the `quantum symmetry' of LG orbifolds and 
the diagonal action of $(\sigma_p,\tilde{\sigma}_p)$ corresponds to an order $p$ orbifold with a 
specific discrete torsion, see~\cite{Israel:2015efa,Hull:2017llx} for details. 

In this work, as in~\cite{Israel:2013wwa,Hull:2017llx}, we focus on freely acting 
orbifolds of the type IIA superstring compactified on $K3 \times T^2$. We supplement the action of the mirrored automorphism $\hat{\sigma}_p$ on the 
$K3$ CFT with an order $p$ translation along a one-cycle of the two-torus. 

The orbifold CFT gives in space-time a four-dimensional theory with $\mathcal{N}=2$ supersymmetry. Unlike compactifications on CY 3-folds, all space-time 
supersymmetry comes from the left-movers, signaling the non-geometric nature of 
the compactification. An important consequence is that, from the point of view of the low-energy 4d theory, the dilaton lies in a vector multiplet and not 
in a hypermultiplet. Furthermore, a large part, if not all, of the $K3$ moduli are lifted, see~\cite{Israel:2013wwa} 
for a detailed analysis of the massless spectra and for   one-loop partition functions of the models.  The moduli spaces of these theories will be analysed in~\cite{ghi}.

To summarize, mirrored automorphisms are non-geometric symmetries of $K3$ CFTs in the Landau-Ginzburg regime, 
and are associated with   isometries of the total cohomology lattice 
$\Gamma_{4,20}$ that have no invariant sublattices. Freely-acting orbifolds constructed from the action of 
a mirrored automorphism on a $K3$ Gepner model together with a translation along 
the two-torus give rise to interesting $\mathcal{N}=2$ non-geometric compactifications of 
type IIA superstrings, providing explicit examples of the general construction outlined in section~\ref{sec:typeII}.
The heterotic duals of these models will be found in the next section.

\section{Heterotic Duals of Non-Geometric Type II Compactifications}
\label{sec:hetdual}

%\subsection*{Heterotic - type IIA dualities}

The  remarkable string theory duality  between the type IIA 
superstring theory compactified on a K3 surface and 
heterotic string  theory compactified on a four-torus~\cite{Hull:1994ys,Witten:1995ex}
 is non-pertur\-bative, 
in the sense that it maps the strongly-coupled regime of the heterotic string 
to the weakly-coupled limit of the type IIA string and \textit{vice versa} (for a review, see~\cite{Aspinwall:1996mn}).
From this fundamental duality one can infer numerous other connections between string theories with lower dimensionality. 

The duality-twisted reduction on a further $T^2$ of the IIA string on K3  reviewed in sections 2 and 3 should then be dual to a
duality-twisted reduction on a further $T^2$ of the heterotic string  on $T^4$.
At a fixed point, the orbifold of the IIA string on $K3\times T^2$ should then be dual to an orbifold of the heterotic string on $T^6$.
On the IIA side, the orbifold is by the symmetry $(\gamma, t)$ where $\gamma $ is a mirrored automorphism of K3 and $t $ is a shift on $T^2$.
On the heterotic side, $O(\Gamma_{4,20})$
 is the heterotic T-duality group, suggesting that the heterotic  dual could be
 the asymmetric orbifold of the heterotic string on $T^6$ by
 $(\gamma, t)$, where $\gamma$ is a heterotic T-duality and $t $ is the same shift on $T^2$ as before.
 However,  duality and orbifolding  do not  necessarily commute in general, so this 
 conjectured dualisation needs further examination.
 
The general idea behind heterotic/type II duality in four dimensions is 
to apply fibre-wise the  duality between a K3 fibration over some base $B$ on the type IIA side 
and  a $T^4$ fibration over $B$ on the heterotic side~\cite{Aspinwall:1995vk}.  Our construction has a base $B=T^2$ and does not constrain the size of the 
$T^2$ part of the (type IIA) internal space, so that one could go to the decompactification 
limit of the $T^2$ base; moreover, the action of the automorphism $(\gamma, t)$ is free, so that the quotient does not 
develop singularities.  Under these two conditions the adiabatic argument of~\cite{Vafa:1995gm} holds and, at least in the limit of a large $T^2$ base 
which allows to perform the dualtiy 'locally' on the fibre,  the heterotic dual should be the asymmetric orbifold of the heterotic string on $T^6$ by  $(\gamma, t)$.

We shall show that this correspondence must be refined for small $T^2$, with heterotic string winding mode contributions modifying the orbifold (this type of 
contribution to heterotic/type II dual pairs was anticipated already in~\cite{Vafa:1995gm}). Specifically, the 
automorphism $(\gamma, t)$ must be supplemented by an order $p$ shift in the T-dual circle conjugate to winding number, so that the full orbifold is by $(\gamma, t  )$
where now $t$ is a shift on both the original $T^2$ and the T-dual $T^2$.
This modification of the heterotic orbifold in turn implies a non-perturbative modification of the type IIA orbifold.

Our construction has some similarities with the model of Ferrara, Harvey, Strominger and Vafa (FHSV)~\cite{Ferrara:1995yx} 
which relates type IIA compactified on the Enriques Calabi-Yau 3-fold, 
which is a freely-acting orbifold of $K3\times T^2$, 
to heterotic strings compactified on a freely-acting, asymmetric orbifold of $T^6$. 
In the FHSV construction, the automorphism acts freely on the $K3$ surface  and 
has fixed points on the base. In our case it is   the opposite: the automorphism 
acts freely on the two-torus and has fixed points on the $K3$ surface. The comparison between these 
two classes of models will be further developed in sections~\ref{sec:mirror} and~\ref{sec:dualmir}.

\subsection{Type IIA - Heterotic Duality in Six Dimensions}

The type IIA  string compactified  on a $K3$ surface gives  $(1,1)$ supergravity in six dimensions at low energies. The moduli space is given by 
\begin{equation}
\label{eq:modulispace}
\mathscr{M}_\textsc{6d} \cong O(\Gamma_{4,20}) \backslash O(4,20) \slash \big( O(4) \times O(20)\big) \times \mathbb{R}\, 
\end{equation}
where the first factor is the moduli space~(\ref{eq:NLSMmoduli}) of non-linear sigma-models on $K3$  and the second is the dilaton zero-mode. 
BPS states arise from branes wrapping cycles of the K3, and include a 
 BPS solitonic string obtained by wrapping an 
NS5-brane around the K3 surface. 

In the duality between  the Type IIA string compactified on K3 and heterotic strings compactified on $T^4$, 
the six-dimensional dilatons and metrics are related by $\phi_\textsc{het}=-\phi_\textsc{iia}$ and 
$g_\textsc{het} = \exp (2 \phi_\textsc{het})\, g_\textsc{iia}$.
 The heterotic 
moduli space is again~(\ref{eq:modulispace}), but now with the first factor being  interpreted as the moduli space 
of Narain lattices of signature $(4,20)$~\cite{Narain:1985jj,Narain:1986am}.
 %{\it i.e.} of  embeddings of the lattice $\Gamma_{4,20}$ of winding and momenta into $\mathbb{R}^{4, 20}$
%it is understood locally as the Grassmannian of space-like 4-planes in $\mathbb{R}^{4, 20}$. 
Geometrically, it describes the moduli space of flat metrics and constant B-fields on $T^4$ and of  $U(1)^{16}$ Wilson lines on $T^4$. While the $O(4)$ factor rotates 
the left-moving bosons of the free CFT with $T^4$ target space, the $O(20)$ factor mixes together the right-moving bosons of the $T^4$ CFT with the 16 bosons describing the gauge sector. 
The duality group $O(\Gamma_{4,20})$ is    the heterotic T-duality group of the heterotic string on $T^4$.
 The second term in~(\ref{eq:modulispace}) is now
  the heterotic dilaton zero-mode.

The heterotic theory has a BPS solitonic string arising from an NS5-brane wrapping the $T^4$; under the duality, it is   mapped to the type IIA 
fundamental string, while   the type IIA NS5-brane wrapped on K3 maps to the heterotic fundamental string~\cite{Harvey:1995rn}. 
Such a correspondance will  be useful in the analysis of BPS states in section~\ref{sec:bps}.

\subsection{Construction of the Heterotic Dual}

The starting point of our construction is a point in the moduli space~(\ref{eq:NLSMmoduli}) that is invariant under a $\mathbb{Z}_p$ automorphism generated by an element $\gamma \in O(\Gamma_{4,20})$. Viewing this as a type IIA construction, this is the type IIA string  compactified on $K3 \times T^2$, where the K3 is chosen to be at a Gepner point in the K3 moduli space so that the corresponding CFT is given by a Gepner model described in~\cite{Hull:2017llx} (\textit{e.g.} the K3 at the Gepner point is~\eqref{eq:23824} in the example given in section~\ref{sec:typeII}).  The automorphism $\gamma$ acts on the K3 as a mirrored automorphism $\hat \sigma $.

In the dual heterotic interpretation, the moduli 
space~(\ref{eq:NLSMmoduli}) is  viewed   as the moduli space of Narain lattices of signature $(4,20)$, acted on by the heterotic T-duality group $O(\Gamma_{4,20})$.
 The special point in moduli space corresponds to a Narain lattice with enhanced discrete symmetry
but without enhanced non-Abelian gauge symmetry.\footnote{For instance, the heterotic   lattice associated with the dual type IIA Gepner model for the 
K3 surface~\eqref{eq:23824} has a discrete symmetry   $[(\mathbb{Z}_2 \times \mathbb{Z}_{3} \times \mathbb{Z}_8 \times \mathbb{Z}_{24})/\mathbb{Z}_{24} ]\times \mathbb{Z}_{24}$.} 

Then
$\gamma$ is an element of the discrete group $O(\Gamma_{4,20})$ that is
$O(4,20)$-conjugate to an element $M$ of the compact subgroup $O(4)\times (20)$, i.e. there is a $U\in O(4,20)$ and $M\in O(4)\times (20)$ so that 
(\ref{eq:gamM}) holds.
The transformation $M$ in $O(4)\times (20)$
 is specified by two angles  characterising a rotation in $O(4)$ and 10 angles characterising a rotation in $O(20)$. 
 As it satisfies $M^p=1$, the angles in $O(4)$ 
are $2\pi s_1 /p ,2\pi s_2/p $ for integers $s_1,s_2$ and the angles in $O(20)$ are
$2\pi r_1/p ,\dots 2\pi  r_{10}/p$ for integers $r_1,\dots , r_{10}$.
An important result from~\cite{Hull:2017llx} is that for a mirrored automorphism  none of the angles is zero, so that no directions are left invariant by the rotation.

Recall from section~\ref{sec:typeII} that at a point in the moduli space~(\ref{eq:NLSMmoduli}) given by the coset representative $(g_0)$, the stabilizer is
\begin{equation}
H_0= O(4)_0\times O(20)_0 \, ,
\end{equation}
where $O(4)_0$ and $O(20)_0$ are the subgroups of $O(4,20)$ defined in eqns.~(\ref{grops_a},\ref{grops_b}). 
An important point is that, in the heterotic string realisation,  at the point $(g_0)$ in the moduli space
$O(4)_0$
 acts only on the left-movers of the heterotic string and $O(20)_{20}$ acts only on the right-movers.
In particular, the vector $\Pi$  at $(g_0)$ encoding the heterotic momenta and winding numbers and taking values in the lattice $\Gamma_{4,20}$
decomposes into a 
4-component momentum $\Pi_L$ with contributions only from left-moving degrees of freedom and transforming under $O(4)_0$ but not  $O(20)_0$
 together with a 
20-component momentum $\Pi_R$ with contributions only from right-moving degrees of freedom and transforming under $O(20)_0$ but not  $O(4)_0$.
Taking  $(g_0)$ to be
 the special point in moduli space, the twist $\gamma$ is in 
 $O(4)_0
\times O(20)_0$.

If $\gamma$ corresponds in the dual type IIA picture to a mirrored automorphism $\hat{\sigma}_p$, a lemma from~\cite{Hull:2017llx} shows that the matrix $M$ representing  it can be diagonalised over the complex numbers to give
\begin{equation}
\label{eq:automorphism_eigenvalues}
M=
	\begin{pmatrix}
		\zeta_p \mathbb{I}_q & 0  & \cdots & \cdots & 0 \\
		0 & \ddots & & & \vdots \\
		\vdots & & \zeta_p^k \mathbb{I}_q & & \vdots \\
		\vdots & & & \ddots & 0 \\
		0 & \cdots & \cdots  & 0 & \zeta_p^{p-1} \mathbb{I}_q
	\end{pmatrix},
\end{equation}
where $\zeta_p$ is a primitive $p$'th root of unity and $k$ takes all values from $1$ to $p-1$ satisfying $\gcd(k,p)=1$; put differently, $\gamma$ has an eigenspace of dimension $q$ for each primitive $p$'th root of unity. The dimension $q$ is therefore equal to $24/\varphi(p)$, where $\varphi(p)$ is the Euler totient function (that is the number of integers $k$ with $k \leq p$ satisfying $\gcd(k,p)=1$). For prime orders $p \in \lbrace 2, 3, 5, 7, 13 \rbrace$, the eigenspaces of $\gamma$ are then all $24/(p-1)$-dimensional.

The type IIA construction is an orbifold of the IIA string on $T^2\times K3$ by the $\mathbb{Z}_p$ symmetry generated by $(\gamma, t)$
 where $\gamma$ is a mirrored automorphism and $t$ is a shift of order $p$ on one of the circles.
On the heterotic side, we have a $\mathbb{Z}_p$ orbifold by the twist $\gamma$ acting as a T-duality automorphism of the Narain lattice together with a shift $t$.
In the large volume limit of the $T^2$ base, the shift should agree with that on the type II side, 
but as discussed above, we will need to consider more general shifts here.
As explained \textit{e.g.} in~\cite{GrootNibbelink:2017usl}, any component of a  shift vector along  directions in which the 
twist acts non-trivially may be absorbed by a redefinition of the origin of  the coordinates,
 so that without loss of generality 
one may consider a shift only along the directions in which $\gamma$ acts trivially.  In other words, decomposing the full Narain lattice of winding and momenta as
\begin{equation}
\Gamma_{6, 22} \cong \Gamma_{4, 20} \oplus \Gamma_{2, 2},
\end{equation}
we quotient by an order $p$ twist $\gamma$ in $O(\Gamma_{4,20})$, so acting non-trivially on the $\Gamma_{4, 20}$ lattice only, 
together with a shift $t$ %along $\Gamma_{2,2}$ 
defined by a lattice vector $\delta$ such that $p \, \delta \in \Gamma_{2, 2}$ but $\delta \notin \Gamma_{2, 2}$.

It is easy to check that $\mathcal{N}=2$ supersymmetry is preserved by the heterotic  orbifold   in this picture. Indeed, the action of $\gamma$ on the world-sheet 
fermions is deduced from its action on the left-moving bosons, as usual, by requiring world-sheet supersymmetry to be preserved. 
The group $SO(4)_0$ acts on the left-handed Ramond ground states as a spinor,
transforming as a $(2,1)+(1,2)$ under $Spin(4)\sim SU(2)\times SU(2)$.
If $s_1=\pm s_2$, the twist is in just one of the two $SU(2)$ subgroups and so half the spinor degrees of freedom remain massless.

\subsection{Geometric and Non-geometric Twists}

The twist $\gamma$ is in  the intersection of the $O(\Gamma_{4,20}) $ and $ O(4)_0\times O(20)_0 $ subgroups of $O(4,20)$.
If the $O(20)_0 $ part of the twist is in fact in an $ O(4)_0\times O(16)_0$ subgroup, then the twist can be regarded as an $ O(4)_0\times O(4)_0\subset O(4,4)$ 
transformation acting on the $T^4$ CFT and an
$O(16)_0$ transformation acting on the gauge degrees of freedom. If moreover it is in the diagonal subgroup $O(4)_{diag} \subset
O(4)_0\times O(4)_0$, then it is a geometric transformation (rotation) on the  $T^4$ and the orbifold is a conventional 
(not asymmetric) orbifold of $T^4$ combined with an (unconventional) orbifold action on the gauge sector. This of course requires choosing an $O(4)$ 
subgroup of $O(20)$, and acting with $O(\Gamma_{4,20}) $ can change such  a \lq geometric' orbifold to a non-geometric one. However, the twist in $O(16)_0$ may 
not be related to the orbifold limit of an ordinary vector bundle.  This will be discussed further in section~\ref{sec:dualmir}.

Bearing this in mind, we now address the question of whether a 
  given theory is dual to a geometric orbifold via an $O(\Gamma_{4,20})$ 
transformation. The answer turns out to depend strongly on the order of the orbifold. For simplicity we discuss only the cases with $p$ prime below. 

In the $p=2$ case, $\Gamma_{4,20}$ is quotiented by the involution which flips all directions of the lattice; therefore, as the twist $\gamma$ 
may be represented here by $-\unit$, it does not mix space-time and gauge degrees of freedom, so that its restriction to the four-torus admits a standard 
geometric interpretation (as the same involution that gives a $T^4/\mathbb{Z}_2$ orbifold). Moreover, $\gamma$ therefore remains the same under 
$O(\Gamma_{4,20})$ conjugation so that the resulting theory always has a geometric interpretation.\footnote{According to~\cite{Vafa:1995gm}, this action 
on the $\Gamma_{4,20}$ lattice is the same as the action of  $(-1)^{F_L}$ on  type IIA compactified on K3.}

In the $p=3$ case, looking for a representative of the conjugacy class of a twist which belongs to $O(4)_{diag} \times O(16)_0$ is not 
straightforward in general. However, it is possible to show that the explicit example of an order 3 twist given in~\cite{Hull:2017llx} 
may be understood as having a geometric action (this may be seen using for instance the parametrisation of $O(\Gamma_{4,20})$ 
of~\cite{GrootNibbelink:2017usl}). Therefore, there exist models in the $p=3$ case which are equivalent to geometric theories 
from the heterotic point of view.

The $p=5$ case is more tricky, as no explicit matrix representation of $\hat{\sigma}_5$ is known  by the authors. 
It is known that there are no supersymmetric $T^4$ orbifolds of order five, see {\it e.g.}~\cite{Walton:1987bu}.  
A simple argument given in~\cite{Gaberdiel:2012um} rules out the possibility of a left-right symmetric action of the orbifold on the $T^4$. 
Let us assume first that there is an order $5$  twist $\gamma$ with a geometric action, 
that is such that $\gamma \in O(4)_{diag} \times O(16)_0$; then, looking at the action on 
$T^4$, $\mathcal{N}=2$ supersymmetry imposes the trace of its matrix representation to be equal to 
$8 \mathrm{cos}\left(\pm \frac{2\pi}{5}\right) = 2(\sqrt{5}-1)$ through the $|s_1| = |s_2|$ condition 
derived in section~\ref{sec:typeII}. This is  incompatible with the requirement that the twist 
belongs to the duality group of the lattice, as this forces in particular the trace of its matrix 
representation to be integer-valued. Therefore, although there exist rank-4 Euclidean lattices 
admitting an order five symmetry, it is not possible to find a twist whose action would admit a geometric interpretation in the $p=5$ case.

In the $p=7$ and $p=13$ cases, the orbifold must be asymmetric by construction. Indeed, such a construction could only be obtained if the twist were acting as an automorphism in $O(4)_{diag}$ (together with an action on the gauge degrees of freedom). 
A result from
lattice theory states that there exist euclidean lattices $\Lambda$ admitting an order $N$ symmetry if and only if $\mathrm{rank}\ \Lambda \geq \varphi(N)$, $\varphi$ being Euler's totient function as before~\cite{vaidyanathaswamy1928integer}. It is then immediate that no rank-4 lattice may admit an order-$p$ symmetry when $p=7$ or 13. The asymmetry of the construction between left- and right-movers on the $T^4$ is even more striking in the $p=13$ case in which there are exactly two angles of absolute value $2 \pi k/13$ for each $k$ between 1 and 6; therefore, the $\mathcal{N}=2$ supersymmetry condition $s_1 = \pm s_2$ ensures us that there may be no angle $r_I$ equal to any of the $s_i$'s, making the asymmetric nature of the model obvious in this case. Once again, we can therefore conclude that this construction does not admit a standard geometric interpretation in the heterotic framework either.

\subsection{Modular Invariance and Restrictions on the Shift Vector}
\label{subsec:hetdual_modinv}

We now turn to the choice of shift vector in    the heterotic orbifold. The twist $\gamma$ is to be accompanied by a   
translation by a  shift vector $\delta$ with $p\delta \in \Gamma_{2,2}$. The choice of group action 
$\mathbb{Z}_p \subset  U(1)^2 \cong T^2$ on the type IIA side of the duality fixes the momentum associated with this shift vector , {\it i.e.} 
the generator of translations along the corresponding one-cycle, but not its action on states with winding number.

It has long been known that 
in order to preserve modular invariance in orbifold models,
there must be  a relation between twists and shift vectors~\cite{Narain:1986qm,Narain:1990mw}. For later convenience, we define the charge vector $\Delta \in \Gamma_{2, 2} \backslash \left(p\, \Gamma_{2, 2} \right)$ 
so that the shift vector $\delta$ satisfies $\Delta = p\, \delta$. 
As discussed in   section~\ref{sec:typeII}, preserving $\mathcal{N}=2$ space-time supersymmetry imposes $s_1 = \pm s_2$. Using this,  it is   possible to show that the necessary and sufficient condition for modular invariance of our theory is given by\footnote{To be precise, there are two additional constraints in the case of even $p$, namely that $s_1+s_2 = 0\ \mathrm{mod}\ 2$ and $(v, \gamma^{p/2} v) = 0\ \mathrm{mod}\ 2$ for all $v \in \Gamma_{6,22}$; however, the first condition must be fulfilled as a consequence of the $\mathcal{N}=2$ supersymmetry preserved by the orbifold. Moreover, as $\gamma^{p/2}$ acts as minus the identity operator on $\Gamma_{4,20}$ and trivially on $\Gamma_{2,2}$ for any even $p$, the second condition is taken care of by the fact that $\Gamma_{6,22}$ is an even lattice.}~\cite{Narain:1986qm,Narain:1990mw}
\begin{subequations}
\label{eqs:modular_invariance}
	\begin{equation}
	\label{eq:modular_invariance_b}
	\Delta^2 + \sum_{I = 1}^{10} r_I^2 = 0 \qquad \mathrm{mod}\ f p \qquad \mathrm{where}\ f=\left\lbrace\begin{array}{l} 1 \ \mathrm{if}\ p \ \mathrm{is\ odd} \\2 \ \mathrm{if}\ p \ \mathrm{is\ even} \end{array}\right.
	\end{equation}
\end{subequations}
Furthermore, the spectrum of $\gamma$ is completely fixed; indeed, equation~\eqref{eq:automorphism_eigenvalues} shows that, of the 12 angles, there are exactly $\frac{12}{\varphi(p)}$ angles equal to $\frac{k}{p}$ mod $1$   for each value of $k$ between 1 and $p-1$ such that $\gcd(k,p)=1$. Then, as shown in appendix~\ref{app:part_function}, the quantity $\sum_{I=1}^{10} r_I^2 + \sum_{i=1}^2 s_i^2$ may be explicitly computed for any $p$, so that equation~\eqref{eq:modular_invariance_b} may be simplified to become
\begin{equation}
  \label{eq:modular_invariance_c}
  \Delta^2 = 2 \Psi_p \qquad \mathrm{mod}\ f p
\end{equation}
where
\begin{equation}
  \label{eq:modular_invariance_c}
\Psi_p :=  %\left[
s^2 - \prod_{\substack{q | p \\ q\, \mathrm{prime}}} (-q) %\right] 
\end{equation}
with  the product running over the distinct prime divisors $q$ of $p$. We parametrise the shift vector as $\delta = (\alpha ^i, \beta _i)$ so that $\Delta = p (\alpha ^i, \beta _i)$ is a lattice vector and the shift is generated by $\alpha ^i k_i + \beta _i w^i$ where $k_i$ and $w^i$ are respectively the momentum and winding charges; the constraint~\eqref{eq:modular_invariance_c} then becomes the condition
\begin{equation}
    \label{eq:modular_invariance_e}
 p^2 \alpha ^i \beta _i = \Psi_p  \mod p \, ,
\end{equation}
which prevents $\beta$ from vanishing, as $\Psi_p$ may never be vanishing modulo $p$ (since any $q$ in the above product is a divisor of $p$ and as such is not invertible over $\mathbb{Z}_p$, unlike $s^2 \in \mathbb{Z}_p^\times$). This translates into  a non-perturbative modification of the orbifold from the type IIA perspective that will be discussed in section~\ref{sec:mirror}. As a side remark, we may note that whenever $p$ is square-free, the condition~\eqref{eq:modular_invariance_c} simplifies to
\begin{equation}
  \label{eq:modular_invariance_prime}
  \Delta^2 = 2 s^2 \  \mathrm{mod}\  f p\, ;
\end{equation}
in particular, equation~\eqref{eq:modular_invariance_prime} holds for $p$ prime. One can further simplify this condition by choosing $s=1$, {\it i.e.} that the rotation in $O(4)_0$ corresponds to the angles $2\pi/p$ and $\pm 2\pi/p$ (any other choice 
is related to this one by relabelling the orbifold sectors), so that 
\begin{equation}
  \label{eq:modular_invariance_prime_simp}
  \Delta^2 = 2 \ \mathrm{mod}\  f p\, .
\end{equation}

Let us now derive the partition function of the theory in order to check explicitly the relations~\eqref{eqs:modular_invariance}. As usual with conformal 
field theories defined on orbifolds, the partition function of the model may be expressed as a sum
\begin{equation}\label{sectorsum}
  Z (\tau,\bar \tau)= \frac{1}{p} \sum\limits_{k,l = 0}^{p-1} Z\jtheta{k}{l}(\tau,\bar \tau)
\end{equation}
 over all allowed boundary conditions (that is, twisted or untwisted), 
with the contribution from the     $ ({k},{l})$ sector defined as
\begin{equation}
  Z \jtheta{k}{l} (\tau,\bar \tau) :=  \mathrm{Tr}_{\mathcal{H}_k} \left(g^l q^{L_0 - \frac{c}{24}} \bar{q}{^{\bar{L}_0 - \frac{\bar{c}}{24}}}\right),
\end{equation}
where $g$ is the element of the point group whose action combines a twist by $\gamma$ and a shift by $t$ and where $\mathrm{Tr}_{\mathcal{H}_k}$ 
stands for the trace over states in the sector twisted by $g^k$.  The various blocks of the partition function are then computed in the usual way to give\footnote{Conventions and properties 
of the Jacobi $\vartheta$-functions are collected in appendix~\ref{app:part_function}.}
\begin{subequations}
  \begin{equation}
    Z\jtheta{0}{0} = \frac{1}{2 \tau_2} \frac{\Theta_{\Gamma_{4,20}} (\tau,\bar \tau) \times \Theta_{\Gamma_{2,2}}(\tau,\bar \tau)}{\eta^{12} (\tau) \bar{\eta}^{24} (\bar \tau)} \sum_{\alpha, \beta = 0}^1 (-1)^{\alpha + \beta + \alpha \beta} \vartheta^4 \jtheta{\alpha}{\beta} (\tau|0)
  \end{equation}
  \begin{multline}
    \label{eq:PartFunct}
    Z\jtheta{k}{l} = \frac{\kappa\jtheta{k}{l}}{2 \tau_2} \exp\Big\{\frac{i \pi k l}{p^2}\Big(2\Psi_p-\Delta^2\Big)\Big\} \frac{\Gamma\jtheta{k}{l}(\tau)}{\left|\eta(\tau)\right|^{12} \bar{F}\jtheta{k}{l}(\bar \tau)} \\
    \times \sum_{\alpha, \beta = 0}^1 (-1)^{\alpha + \beta + \alpha \beta} \vartheta^2 \jtheta{\alpha}{\beta} (\tau|0) \prod_{i=1}^2 \frac{\vartheta\jtheta{\alpha+2 k s_i/p}{\beta + 2 l s_i/p} (\tau|0) \times \bar{\vartheta}\jtheta{1+2 k s_i/p}{1 + 2 l s_i/p} (\tau|0)}{\vartheta\jtheta{1+2 k s_i/p}{1 + 2 l s_i/p} (\tau|0)}\, ,
  \end{multline}
\end{subequations}
where the second equation is only valid for $(k,l) \neq (0,0)$. In the above equations, we have defined $\Theta_\Lambda$ as the sum over charges lying in the lattice $\Lambda$, the sum over the lattice $\Gamma\jtheta{k}{l}$ as
\begin{equation}
  \label{eq:sum_charge_lattice_def}
  \Gamma\jtheta{k}{l}(\tau) := \sum_{Q \in \Gamma_{2, 2} + k \delta} q^{\, Q_L^2/2} \bar{q}^{\, Q_R^2/2} e^{2 i \pi \left<Q,\delta\right>} ,
\end{equation}
the function $F\jtheta{k}{l}$ as
\begin{equation}
  F\jtheta{k}{l}(\tau) := \frac{1}{\eta^{12}(\tau)} \prod_{I=1}^{10} \vartheta\jtheta{1+2 k r_I/p}{1+2 l r_I/p}(\tau | 0) \prod_{i=1}^{2} \vartheta\jtheta{1+2 k s_i/p}{1+2 l s_i/p} (\tau | 0)
\end{equation}
and the ``degeneracy factor'' $\kappa\jtheta{k}{l}$ as
\begin{equation}
\kappa\jtheta{k}{l} := \prod_{d | p} \left(\frac{d}{\gcd(k,l,d)}\right)^{\frac{12}{\varphi(p)} \gcd(k,l,d) \mu\left(\frac{p}{d}\right)} \, .
\end{equation}
One may notice that the phase factor in the partition function block~\eqref{eq:PartFunct} may be set to one  by choosing an appropriate representative of the shift vector $\delta \in \frac{1}{p} \Gamma_{2,2} / \Gamma_{2,2}$, as shown in appendix~\ref{app:part_function}. As discussed in the introduction, the Narain lattice $\Gamma_{4,20}$ appearing in the $(0,0)$ sector lies at a point in moduli space corresponding, on the type IIA side, to a Gepner model admitting a mirrored automorphism of order $p$.

Anticipating the following section, we    emphasise here that no sum over the charge lattice $\Gamma_{4,20}$ appears (except for the term with $(k,l)=(0,0)$), which is due to the fact that the twist $\gamma$ acts non trivially on the whole lattice; hence any state with non-vanishing momentum in $\Gamma_{4, 20}$ is projected out in the orbifolding procedure 
  (see \textit{e.g.}~\cite{Narain:1986qm} for an extensive discussion). On the type IIA side of the duality, it means that there is no lattice of BPS $D$-brane charges, which is easy to understand as that theory has no massless Ramond-Ramond ground states~\cite{Israel:2013wwa}. There is no non-abelian gauge group enhancement coming from $\Gamma_{4,20}$ on the heterotic theory
as we are at a point in the moduli space corresponding to a non-singular K3 CFT.

The full partition function of the heterotic orbifold CFT, given by the sum (\ref{sectorsum}) over all  sectors,
is therefore modular invariant provided that 
\begin{equation}
Z\jtheta{k + p}{l} = Z\jtheta{k}{l+p} = Z\jtheta{k}{l} \ , \quad \forall\,  k,l\, .
\end{equation}
This is precisely what is ensured by the equations~\eqref{eqs:modular_invariance} which are therefore interpreted, in the heterotic picture, as necessary constraints on the shift vector to obtain a (perturbatively) well-defined string vacuum. In short, a non-vanishing winding shift is imposed by the modular invariance constraints.

\section{BPS States}
\label{sec:bps}

 BPS states  have dual interpretations in the two dual theories,
the type IIA theory on $K3\times T^2$ and the heterotic string on $T^4\times T^2$ ~\cite{Hull:1994ys} (see \textit{e.g.} 
Table 1 of~\cite{Dabholkar:2005dt}).  
 In particular, winding and momenta along one-cycles of the four-torus in the heterotic theory correspond to D-branes 
wrapping     cycles of K3 in the type IIA description of the theory, while  
momentum and winding states on $T^2$ in the heterotic picture are respectively 
understood as momentum states on $T^2$ and as NS5 branes wrapping $K3 \times S^1 \subset K3 \times T^2$ from the type IIA perspective. 
On the type IIA side, after the quotient by the mirrored K3
automorphism, no D-brane states remain; this is due to the fact that
space-time supersymmetry is entirely carried by left-movers so that there
are no massless Ramond-Ramond p-forms hence no BPS Dp-branes.
In the heterotic dual, this
corresponds
to the fact that
 fundamental strings with momentum and/or winding   on the 4-torus
 are projected out, as the  automorphism  used in the quotient leaves no cycle of $T^4$ invariant. 
Fundamental heterotic  strings with winding around a one-cycle of the $T^2$ 
are dual to 
 the type IIA NS5-brane wrapping the same one-cycle of the base together with 
the $K3$ fibre; on taking the quotient, this descends to what can be thought of as an NS5-brane wrapping a \lq cycle'
 of the non-geometric Calabi-Yau background.\footnote{For a discussion of branes in non-geometric backgrounds, see e.g. \cite{Hull:2019iuy} and references therein.}

In this section, we shall study the BPS states that arise in the perturbative spectrum of the heterotic string orbifold. The type IIA duals of these states will in general be non-perturbative states 
carrying NS5-brane charge.

\subsection{Helicity Supertraces}

In practice, a powerful tool in studying BPS states is the computation of helicity supertraces, that are protected quantities which do not change when the string coupling is increased; 
however, in four-dimensional theories with $\mathcal{N}=2$ supersymmetry, they can 
jump across walls of marginal stability in the moduli space.  It can be shown (see \textit{e.g.}~\cite{Kiritsis:1997hj} for a review of helicity supertraces properties and references therein) 
that in $\mathcal{N}=2$ theories the only non-vanishing helicity supertrace is
\begin{equation}
  \Omega_2(\mathfrak{R}) := \mathrm{Tr}_\mathfrak{R} \left[(-1)^{2 J_3} J_3^2\right],
\end{equation}
for any representation $\mathfrak{R}$ of the $\mathcal{N}=2$ algebra, with $J_3$ the space-time helicity operator. $\Omega_2$ vanishes for any (long) massive representation of 
$\mathcal{N}=2$ supersymmetry while it is unchanged under recombinations of two BPS multiplets into a long multiplet or \textit{vice versa}, 
making it a well-defined quantity on the moduli space.

In the heterotic frame, it will receive contributions from the perturbative Dabholkar-Harvey (DH) half-supersymmetric BPS states~\cite{Dabholkar:1989jt} that are heterotic fundamental strings in their 
left-moving superconformal ground state characterized by their winding and momentum charges on the torus. It is possible to extract $\Omega_2$ from the partition function 
by introducing a chemical potential for the helicity; more precisely, defining
\begin{equation}
  Z(\tau, \bar \tau |v, \bar{v}) := \mathrm{Tr}_\mathcal{H} \left[(-1)^{2 J_3} e^{2 i \pi v J_3^{(L)} + 2 i \pi \bar{v} J_3^{(R)}} q^{L_0} \bar{q}^{\bar{L}_0}\right],
\end{equation}
with $\mathrm{Tr}_\mathcal{H}$ the trace over the whole Hilbert space of the theory and with $J_3^{(L)}$ and $J_3^{(R)}$ the left and right moving components of the helicity respectively, $\Omega_2$ is generated by the function $B_2$ defined as
\begin{equation}
  \label{eq:BPS_index_def}
  B_2 (\tau,\bar \tau):= \left.\left(\frac{1}{2 i \pi} \frac{\partial}{\partial v} + \frac{1}{2 i \pi} \frac{\partial}{\partial \bar{v}} \right)^2 Z(\tau, \bar \tau |v, \bar{v})
\right|_{v=\bar{v}=0} = \sum_{Q \in \Lambda} \Omega_2(Q) q^{L_0} \bar{q}^{\bar{L}_0}.
\end{equation}
where $\Lambda$ stands for the lattice of electric charges of the orbifolded theory, given here by
\begin{equation}
  \Lambda = \bigoplus_{k=0}^{p-1} \left(\Gamma_{2,2} + k \delta\right) .
\end{equation}

Using the results from section~\ref{sec:hetdual} and identities from appendix~\ref{app:part_function}, one may then show that the modified partition function reads

\begin{equation}
  Z (\tau, \bar \tau |v, \bar{v}) = \frac{1}{p} \sum_{k,l = 0}^{p-1} Z\jtheta{k}{l}(\tau, \bar \tau |v, \bar{v}) ,
		\end{equation}
		where the orbifold blocks are now  
		\begin{subequations}
		\begin{equation}
		Z\jtheta{0}{0} (\tau, \bar \tau |v, \bar{v}) = \frac{1}{\tau_2} \frac{\Theta_{\Gamma_{4,20}} (\tau,\bar \tau) \times \Theta_{\Gamma_{2,2}}(\tau,\bar \tau)}{\eta^{12} (\tau) \bar{\eta}^{24} (\bar \tau)} \xi(\tau | v) \bar{\xi}(\bar \tau | \bar v) \vartheta_1^4 \left(\tau|\frac{v}{2}\right)
		\end{equation}
		\begin{multline}
		Z\jtheta{k}{l} (\tau, \bar \tau |v, \bar{v}) = \frac{\kappa\jtheta{k}{l}}{\tau_2} \exp\left[\frac{i \pi k l}{p^2}\left(2\Psi_p-\Delta^2\right)\right] \frac{\Gamma\jtheta{k}{l}(\tau)}{\left|\eta(\tau)\right|^{12} \bar{F}\jtheta{k}{l}(\bar \tau)} \\
    \times \xi(\tau | v) \bar \xi(\bar \tau | \bar v) \vartheta_1^2\left(\tau | \frac{v}{2}\right) \prod_{i=1}^2 \frac{\vartheta\jtheta{1+2 k s_i/p}{1 + 2 l s_i/p} \left(\tau|\frac{v}{2}\right) \times \bar{\vartheta}\jtheta{1+2 k s_i/p}{1 + 2 l s_i/p} (\tau|0)}{\vartheta\jtheta{1+2 k s_i/p}{1 + 2 l s_i/p} (\tau|0)}\, .
  \end{multline}
	\end{subequations}
Here, $\xi(\tau|v)$ is the usual space-time transverse bosons helicity generating function defined as
\begin{equation}
  \xi(\tau|v) := \prod_{n=1}^\infty \frac{(1-q^n)^2}{(1-q^n e^{2 i \pi v})(1-q^n e^{-2 i \pi v})}.
\end{equation}
Differentiating $Z(v,\bar{v})$ with respect to $v$ and $\bar{v}$ gives the index $B_2$:\begin{multline}
  \label{eq:BPS_index}
  B_2 = -\frac{1}{2 p \tau_2} \sideset{}{'}\sum_{k,l} \exp\left[\frac{i \pi k l}{p^2}\left(2 \Psi_p-\Delta^2\right)\right] \frac{\kappa\jtheta{k}{l} \Gamma\jtheta{k}{l}(\tau)}{\bar \eta^6(\bar \tau) \bar{F}\jtheta{k}{l}(\bar \tau)} \prod_{i=1}^2 \bar{\vartheta}\jtheta{1+2 k s_i/p}{1 + 2 l s_i/p} (\tau|0)
\end{multline}
where the primed sum $\sum\nolimits'_{k,l}$ stands for the sum running over all values of $(k,l)$ in $\mathbb{Z}_p \times \mathbb{Z}_p$ except $(0,0)$. 
Note that the term with $(k,l)=(0,0)$, {\it i.e.} the untwisted sector contribution with no quotienting group element insertion,  does not contribute to the index. This illustrates once more the absence, in the orbifolded theory, of states with charges lying in the $\Gamma_{4,20}$ lattice.

As the automorphism generating the $\mathbb{Z}_p$ group we are quotienting by
  has a non-trivial action on the charge lattice, one cannot factorise the BPS index as the product of a sum over the charge lattice by a function with well-defined modular properties; however, it is still possible to split it into smaller blocks which factorise in a similar way by expressing the charge lattice as
\begin{equation}
  \Lambda = \bigoplus_{k,a=0}^{p-1} \Lambda^{(k,a)} ,
\end{equation}
where we define $\Lambda^{(k,a)}$ as
\begin{equation}
  \Lambda^{(k,a)} := \left\lbrace v + k \delta \in \Gamma_{2,2} + k \delta \middle| \left<v,\delta \right> = \frac{a}{p}\ \mathrm{mod}\ 1 \right\rbrace.
\end{equation}
Each $\Gamma\jtheta{k}{l}$ may then be expressed in terms of the   theta functions associated with the lattices $\Lambda^{(k,a)}$, for $a$ between 0 and $p-1$. This allows one to extract from the BPS index~\eqref{eq:BPS_index} the indices for each sublattice $\Lambda^{(k,a)}$ of the charge lattice, as all charges in a given $\Lambda^{(k,a)}$ transform in the same way under the whole automorphism $g_p$. Defining as before $\Theta_{(k,a)}$ as
\begin{equation}
  \Theta_{(k,a)} := \sum_{Q \in \Lambda^{(k,a)}} q^{Q_L^2/2} \bar{q}^{Q_R^2/2} ,
\end{equation}
the whole $B_2$ index may be expressed as $B_2 = \sum_{k,a = 0}^{p-1} B_2^{(k,a)} \Theta_{(k,a)}$ with
\begin{multline}
  \label{eq:split_B2}
  B_2^{(k,a)} := -\frac{1}{2 p \tau_2} \sum_{l=0}^{p-1} \exp\left[\frac{i \pi k l}{p^2}\left(2\Psi_p+\Delta^2\right) + \frac{2 i \pi a l}{p}\right] \\
  \times \frac{\kappa\jtheta{k}{l}}{\bar\eta^6(\bar\tau) \bar{F}\jtheta{k}{l}(\bar \tau)} \prod_{i=1}^2 \bar{\vartheta}\jtheta{1+2 k s_i/p}{1 + 2 l s_i/p} (\bar \tau|0)
\end{multline}

There is a subtlety to take into account here; the definition~\eqref{eq:BPS_index_def} of $B_2$ implies that two different charges $Q$ and $P$ will contribute to the same index 
$\Omega_2$ if they satisfy $Q_i^2 = P_i^2$ for $i = L, R$ ($Q_L$ and $Q_R$ standing for the left and right components of the charge vector $Q$ respectively, as before). 
In particular, this means that opposite charge vectors $Q$ and $-Q$ always contribute to the same index $\Omega_2(Q)$; from a   more physical point of view, this is 
a reflection of the CPT invariance of the theory which imposes that any representation of the $\mathcal{N}=2$ supersymmetry algebra must be accompanied by its CPT 
conjugate, which has charge $-Q$, to form a CPT-invariant multiplet. This   means that the index $\Omega_2(Q)$ must be computed by taking into account not only the 
contributions from $B_2^{(k,a)}$ but also from the possible non-trivial degeneracies of states in the sum over the charge lattice.

\subsection{Some Explicit Results}

A straightforward check of the validity of the above indices may be obtained by evaluating the constant term in $B_2^{(0,0)}$; indeed, this will give the index of the $\mathcal{N}=2$ 
supersymmetry multiplets whose charge $Q$ has vanishing norm. At a generic point in moduli space, these are the only massless multiplets of the theory so that this gives some 
insight about the dimension of the Coulomb and Higgs branches. More precisely, one may show (see \textit{e.g.}~\cite{Kiritsis:1997hj}) that the supergravity and vector 
multiplets each contribute   $+1$ to $B_2$ while hypermultiplets each contribute   $-1$.  The classical vector moduli space was shown to be that of the STU model in~\cite{Hull:2017llx}.

With three vector multiplets and one supergravity multiplet, one expects the constant term to be $4-n_H$, with $n_H$ the number of hypermultiplets remaining in the orbifold theory. An explicit expansion of 
$B_2^{(0,0)}$ in power series yields results that match the analysis of the moduli space that will be given in~\cite{ghi}:  one finds for instance respectively $20, 10, 4, 2$ and $0$ massless 
hypermultiplets in the $p=$ 2, 3, 5, 7 and 13 theories.\footnote{The numbers  of hypermultiplets $20, 10, 4, 2$ and $0$
are the quaternionic dimensions of the corresponding hypermultiplet moduli spaces.}

For each specific value of $Q$, it is also possible to extract the index $\Omega_2(Q)$ from the formul\ae$\ $given above. 
Let us consider for simplicity the five-dimensional theory one would get from an orbifold of $T^4 \times S^1$, 
a decompactification limit of the case we have studied so far. The charges may then be parametrised as $Q=(n,w)$, $n$ 
and $w$ being the momentum and winding numbers of the string respectively. Finding $\Omega_2(Q)$ may easily be done by identifying in which sublattice $\Lambda^{(k,a)}$ $Q$ lies and using the level-matching condition
\begin{equation}
  \label{eq:level_matching}
  \frac{Q^2}{2} = N+\alpha_k ,
\end{equation}
where $N$ is the level of the BPS state and $\alpha_k$ 
arises from
  the difference i ground state energy between left- and right-movers. Setting $|s_i|=1\ \mathrm{mod}\ p$ for $i=1,2$ (which amounts to choosing a  generator of the cyclic group $\gamma\in \mathbb{Z}_p   $), $\alpha_k$ may be explicitly computed and is
\begin{equation}
  \alpha_k = \frac{k^2}{p^2}-\frac{k}{p}-\left(\frac{\gcd(k,p)}{p}\right)^2 \prod_{\substack{q | p \\ q\, \mathrm{prime}}}(-q)+\left\lbrace\begin{array}{l l} \frac{1}{2}-\frac{k}{p} & \mathrm{if}\ 0 \leq k\leq \frac{p}{2} \\ \frac{k}{p}-\frac{1}{2} & \mathrm{if}\ \frac{p}{2} \leq k \leq p-1 \end{array}\right.
\end{equation}
for $1 \leq k \leq p-1$ and, of course, $\alpha_0 = -1$ as usual.

Let us take a simple example, say $Q=(1,5)$ in the above notation, and consider also a model with $p=3$; then, setting once again $\delta=(1/3,1/3)$, one has $\left<Q,\delta\right> = 0\ \mathrm{mod}\ 1$ which indicates that $Q \in \Lambda^{(0,0)}$ with the above notations. Now, as $\frac{Q^2}{2} = 5$, $\Omega_2(Q)$ is simply given by the coefficient of the $\bar{q}^5$ term in the power expansion of $B_2^{(0,0)}$ ; computing the first terms in this expansion gives in this specific case $\Omega_2\left[Q=(1,5)\right]=176$. One should remember here that $\Omega_2$ does not represent a degeneracy, {\it per se}, 
as contributions from integer and half-integers spin multiplets are counted respectively positively and negatively\footnote{Here, the ``spin of a multiplet'' is understood to be the spin of the middle state of the multiplet. This make sense as we are only considering here short multiplets, since any long multiplet has vanishing contribution to $B_2$ as explained earlier.}. This explains for instance that for other values of $Q$, one may find negative values of $\Omega_2$ (\textit{e.g.} $\Omega_2\left[Q=(2,2)\right] = -90$).

One may also consider BPS states lying in twisted sectors, which have non-integer charges in general; explicit computations show that $\left|\Omega_2\right|$ seems to grow faster with the 
level $N$ in the twisted sectors than in the untwisted one (\textit{e.g.} $\Omega_2\left[Q=\left(\frac{1}{3},\frac{10}{3}\right)\right] = -236196$, while the two untwisted-sector examples considered above had higher values of $N$ but lower values of $\Omega_2$.). In~\cite{Dabholkar:2005dt}, it was noted that $\Omega_2$ is generically exponentially smaller in  untwisted sectors than in  twisted ones in $\mathcal{N}=2$ orbifold models; explicit expansions of the various $B_2^{(k,a)}$ in powers of $\bar{q}$ seem to confirm this statement.

The asymptotic behaviour of $\Omega_2(Q)$ is also accessible for high values of $Q^2$ following the procedure described in~\cite{Dabholkar:2005dt} which we will briefly review here. As illustrated above with a few examples, the power expansion of the $B_2^{(k,a)}$'s gives us access to the $\Omega_2$ indices; more precisely, writing $B_2^{(k,a)}$ as
\begin{equation}
  B_2^{(k,a)}(\bar{q}) := \sum_N a_N^{(k,a)} \bar{q}^{N-\alpha_k} ,
\end{equation}
it is clear from~\eqref{eq:split_B2} and from the level-matching condition~\eqref{eq:level_matching} that $\Omega_2(Q) = \epsilon(k,a) a_{Q^2/2}^{(k,a)}$ for $Q \in \Lambda^{(k,a)}$, where $\epsilon(k,a)$ is a factor taking into account the fact that both $Q \in \Lambda^{(k,a)}$ and $-Q \in \Lambda^{(-k,-a)}$ contribute to $\Omega_2(Q)$ (explicitely, $\epsilon(k,a) = 1$ if $\Lambda^{(-k,-a)} = \Lambda^{(k,a)}$ and 2 else). Performing an inverse Laplace transform, it is possible to compute $a_N^{(k,a)}$ to find
\begin{equation}
  \label{eq:BPS_asymptotics}
  a_N^{(k,a)} = \int_{\epsilon-i\pi}^{\epsilon+i\pi} \frac{\dr t}{2 i \pi} e^{N t} B_2^{(k,a)}\left(e^{-t}\right) .
\end{equation}
When $N$ reaches high values, the imaginary exponential in~\eqref{eq:BPS_asymptotics} becomes rapidly oscillating so that the integral is dominated by the behaviour of the integrand around $t \sim \epsilon$. In this regime, the $\epsilon \rightarrow 0$ limit of this integral is dominated by a function of $e^{-t} \sim 1$ so that the power expansion of $B_2^{(k,a)}$ is not  useful here; however, one may use the modular properties of $B_2^{(k,a)}$ to replace it by a function of $e^{-4\pi^2/t}$ which is a small parameter when $t$ goes to 0. The above integral becomes

\begin{equation}
  a_N^{(k,a)} = \int_{\epsilon-i\pi}^{\epsilon+i\pi} \frac{\dr t}{2 i \pi} e^{N t} \left(S\cdot B_2^{(k,a)}\right)\left(e^{-\frac{4\pi^2}{t}}\right),
\end{equation}
so that expanding $S\cdot B_2^{(k,a)}$ instead in powers of $\bar{q}$ lead to an approximation of the asymptotic behaviour of $\Omega_2(Q)$ for large values of $\frac{Q^2}{2}$. Here, $S$ is the usual generator of $SL(2,\mathbb{Z})$ acting on the world-sheet parameter $\tau$ as $\tau \mapsto -1/\tau$.

We consider below the models with $p$ a prime number. 
Explicit computations to leading order show that the asymptotic behaviour of $\Omega_2$ in the untwisted sector is given, up to a multiplicative constant, by
\begin{equation}
  \label{eq:BPS_asymptotics_untwisted}
  \Omega_2^\mathrm{untw}(Q) \underset{Q^2 \gg 1}{\sim} \left\lbrace \begin{array}{l c c}
    - \sqrt{\frac{Q^2}{2}} J_1\left(\frac{4 \pi}{3} \sqrt{\frac{Q^2}{2}}\right) & \qquad & p=3 \\
    - \sqrt{\frac{Q^2}{2}} I_1\left(\frac{4 \pi}{5} \sqrt{\frac{Q^2}{2}}\right) & \qquad & p=5 \\
    - \sqrt{\frac{Q^2}{2}} I_1\left(\frac{4 \pi \sqrt{5}}{7} \sqrt{\frac{Q^2}{2}}\right) & \qquad & p=7 \\
    - \sqrt{\frac{Q^2}{2}} I_1\left(\frac{4 \pi \sqrt{29}}{13} \sqrt{\frac{Q^2}{2}}\right) & \qquad & p=13 \\
                                                   \end{array} \right.
\end{equation}
%for any value of $a$.  
Here, $J_1$ ($I_1$) is the (modified) Bessel function of first kind. 
In the twisted sectors, the asymptotic behaviour of the BPS index is surprisingly identical 
%not only for any value of $k$ and $a$ but also 
for any order prime $p$ of the quotienting group; it is then given in all these cases by

\begin{equation}
  \label{eq:BPS_asymptotics_twisted}
  \Omega_2^\mathrm{tw}(Q) \underset{Q^2 \gg 1}{\sim} - \sqrt{\frac{Q^2}{2}} I_1\left(4 \pi \sqrt{\frac{Q^2}{2}}\right) .
\end{equation}
Replacing the Bessel functions by their asymptotic expansions in equations~\eqref{eq:BPS_asymptotics_untwisted} 
and~\eqref{eq:BPS_asymptotics_twisted} then confirms the exponentially small growth of $\left|\Omega_2(Q)\right|$ 
in the untwisted sector compared to that of $\left|\Omega_2(Q)\right|$ in the twisted ones discussed in~\cite{Dabholkar:2005dt}.

\section{The Non-Perturbative Type IIA Construction and Duality}
\label{sec:mirror}

In section~\ref{sec:hetdual}  modular-invariance constraints on the heterotic duals of the type IIA non-geometric Calabi-Yau backgrounds were   analyzed. It was 
 found that perturbative consistency of the heterotic constructions leads to the constraint~(\ref{eq:modular_invariance_e}) 
on the shift vector for the two-torus  and this implies the  shift vector should have non-vanishing winding charge. 
In this section we will examine the consequences of this condition on the type IIA side of the duality, where it leads to a non-perturbative modification 
of the $K3\times T^2$ orbifold. For clarity of the presentation, we will restrict ourselves here to the case in which
 the order $p$ of the automorphism is a prime number.

\subsection{Interpretation of the Shift Vector}

For both the type IIA and heterotic constructions, we have an orbifold by a twist $\gamma \in O(\Gamma_{4,20})$ 
and a shift $t$ on the two-torus by a vector $\delta = (\alpha ^i, \beta _i)$ with $p\, \delta \in   \Gamma_{2,2}$.
The momentum vector $k_i$ ($i=1,2$) on the 2-torus combines with the string winding charges $w^i$ to 
form a generalised momentum vector $\Pi_I=(k_i,w^i) \in  \Gamma_{2,2}$. The shift acts on a momentum state 
$\vert \Pi \rangle= \vert k,w \rangle$ with $k_i, w^i \in \mathbb{Z}$ 
as:
\begin{equation}
\vert \Pi \rangle \mapsto 
\exp ( 2\pi i \delta ^I \Pi _I) \vert \Pi \rangle
\end{equation}
so that
\begin{equation}
\vert k,w \rangle \mapsto \exp (2\pi  i [ \alpha ^i k_i +  \beta _i w^i]) 
 \vert k,w \rangle
\end{equation}
For a shift symmetry of order $p$, {\it i.e.} isomorphic to $\mathbb{Z}_p$, we take 
$\Delta = p \, \delta \in \Gamma_{2,2} $ to be a lattice vector, with norm
\begin{equation}
\Delta ^2 = p^2 \delta ^2 = 2 p^2 \alpha ^i \beta _i \, . 
\end{equation}

If the momenta $k_i$ are realised on the periodic coordinates $y^i \sim y^i+2\pi$ of the 2-torus in the usual way, 
$\exp (2\pi  i \alpha ^i k_i )$ generates the shift
\begin{equation}
y^i \mapsto y^i+2\pi \alpha ^i
\end{equation}
If dual coordinates $\tilde y_i $ conjugate to the winding charge are introduced 
%so that
%\begin{equation}
%w^i = -i \frac {\partial }{\partial \tilde y_i}
%\end{equation}
then $
\exp (   2\pi i \beta _i w^i) $
generates the dual shift
\begin{equation}
\tilde y_i \mapsto \tilde y_i +2\pi \beta _i
\end{equation}
so the shift acts on the coordinates $Y^I=(y^i,\tilde y_i )$ of the doubled torus as 
\begin{equation}
Y^I \mapsto Y^I +2\pi\delta ^I \,.
\end{equation}

In both type IIA and heterotic constructions with a single twist, we can take the  shift to be on a single cycle of the 2-torus, so that $p \, \delta \in 
\Gamma_{1,1} \subset \Gamma_{2,2}$.
In the perturbative type IIA construction, we had 
\begin{equation}
\label{IIAshift}
\delta = (\alpha^1, 0,0,0), \qquad \alpha^1 = \frac 1 p \, ,
\end{equation}
giving a shift
\begin{equation}
y^1\mapsto y^1+ \frac {2 \pi} p \, .
\end{equation}

For the heterotic string, the  modular invariance constraint 
$$\Delta ^2 = 2 p^2 \alpha ^i \beta _i=2 \mod p$$
 obtained in section~\ref{sec:hetdual}, eqn.~(\ref{eq:modular_invariance_prime_simp}),
implies that both $\alpha $ and $\beta $ are non-zero. Setting $\alpha^1 = 1/p$ (in order to match with the perturbative type IIA construction in the 
large $T^2$ limit) one can solve this constraint with 
\begin{equation}
\label{eq:hetshift}
\delta = (\alpha^1, 0,\beta_1,0), \qquad \alpha^1 = \frac 1 p, \qquad \beta_1 =\frac  {1} p \, .
\end{equation}
This vector generates the shifts
\begin{subequations}
\label{dushif}
\begin{align}
y^1 &\mapsto y^1+ \frac {2 \pi} p \, ,\\ 
\tilde y_1 &\mapsto \tilde y_1+ \frac {2 \pi  } p \, .
\end{align}
\end{subequations}
It was to be expected that the shift in $y^1$ should agree in the two pictures, but we see that there is a surprising difference in that   a shift in the dual coordinate $\tilde y_1$
is essential for heterotic modular invariance but there was no corresponding shift on the type IIA side in our construction.

In   our models, the perturbatively consistent type IIA construction determines the heterotic dual in the large volume limit of the $T^2$, with an orbifold by a  twist $\gamma \in O(\Gamma_{4.20})$ and a shift of the coordinate $y^1$ of a cycle of $T^2$. However, away from the decompactification limit 
this heterotic construction is not perturbatively consistent and must be modified by winding number shifts.
Then duality  implies that there should be a dual modification of the type IIA theory. This modification is non-perturbative in the type IIA theory, so does not affect the 
perturbative consistency of the original construction. This is in accord with the discussion of~\cite{Vafa:1995gm}, where it is argued that duality does not   completely determine
the shift vector, and consistency conditions, such as level matching and modular invariance are needed to fix the shift vector. 

A similar situation was encountered in the FHSV model~\cite{Ferrara:1995yx}. We will discuss further this example in  subsection~\ref{subsec:fhsv}, 
and compare it with our models.
In both cases, it is natural to speculate that the modifications in the type IIA theory could arise from a condition for non-perturbative consistency of the IIA string.

\subsection{The Non-Perturbative Type IIA Construction}

A convenient way of representing the modifications to the type IIA construction is as follows.
The transformation $t$ acts on a heterotic state with momentum $k$ and winding $w$ by
\begin{equation}
\label{phasey}
\vert k,w \rangle \mapsto \exp (2\pi  i [ \alpha ^i k_i +  \beta _i w^i]) 
 \vert k,w \rangle
\end{equation}
The type IIA dual of the heterotic momenta $k_i$ and winding charges $w^i$ are some charges $x_i$ and $z^i$ 
in the $\Gamma_{6,22}$ lattice of type IIA compactified on $K3 \times T^2$. For our construction,  $x_i$ remains the  momentum on the torus, so $k_i=x_i$, and 
$z^i$ is the winding charge on the $i$'th circle for the solitonic string obtained by wrapping the IIA NS5-brane on K3, so that $z^i$ 
is the NS5-brane charge for NS5-branes wrapping $K3\times S^1$, with the $S^1$ being the $i$'th circle. (For the FHSV model, 
$x_1$ and $z^1$ are D0-brane and D4-brane charges, as we will discuss in the next subsection.)

Then for the models considered here, the heterotic transformation~(\ref{phasey}) becomes the type IIA  transformation 
\begin{equation}
\label{zphasey}
\vert k,z \rangle \mapsto \exp (2\pi  i [ \alpha ^i k_i +  \beta _i z^i]) 
 \vert k,z  \rangle
\end{equation}
where $k_i$ is the momentum on the $i$'th circle and $z^i$ is the winding number of the solitonic string (from the NS5-brane wrapped on $K3$)
on the $i$'th circle. From eq.~(\ref{eq:hetshift}), consistency of the heterotic perturbative limit is satisfied with 
$\alpha^1=\beta_1 = 1/p$ an $\alpha^2=\beta_2=0$. 

Non-perturbative type IIA states with non-zero winding number for the solitonic string around the first circle of $T^2$ are therefore charged under 
the symmetry used to obtain the non-geometric Calabi-Yau background.  For perturbative states with $z=0$, the transformation~(\ref{zphasey}) 
is of course the same as the one used in the perturbative construction with shift vector~(\ref{IIAshift}).

As we have seen, the action of $t$ on a heterotic state $\vert k,w \rangle$
given by
(\ref{phasey}) gives a shift of the coordinates $y^i$ conjugate to $k_i$ together with a shift of the dual coordinates $\tilde y _i $ conjugate to $w^i$.
Similarly, for the IIA string, if we introduce  coordinates $\hat y _i$ conjugate to $z^i$, then
the action of $t$ on a type IIA state
(\ref{zphasey}) can be understood as  a shift of the coordinates $y^i, \hat y _i$. In general, phase rotations of this kind dependent on brane charges can be reinterpreted as shifts of suitable dual coordinates, justifying our referring to $t$ as a shift; this will be discussed further in the next section.

The above discussion implies, using heterotic/type IIA duality, that  non-perturbatively consistent  non-geometric Calabi-Yau backgrounds 
in type IIA superstring theory should be defined using a shift symmetry of the form~(\ref{zphasey}) that includes a non-perturbative contribution. 
In the FHSV construction that we will discuss below, a similar type of non-perturbative modification of the shift symmetry occurs,   
involving D-brane charges rather than NS5-brane charges.

\subsection{The  FHSV Model}
\label{subsec:fhsv}

The starting point for the FHSV construction~\cite{Ferrara:1995yx} 
is a special K3 surface admitting a freely acting $\mathbb{Z}_2$ involution, such 
that the quotient of $K3$  by this is an Enriques surface. This non-symplectic K3 automorphism acts on the lattice~(\ref{eq:totalcoh}) of total K3 cohomology 
by interchanging two $E_8 \oplus U$ sublattices, acting as $-1$ on one sublattice $U$ and leaving the final $U$ invariant.\footnote{This 
involution is a geometric automorphism, i.e. a large  diffeomorphism of K3, whose action is an element of $O(\Gamma_{3,19})$. The invariant sublattice $U$ is the lattice generated by 
$H^0 (K3;\mathbb{Z})$ and  $H^4 (K3;\mathbb{Z})$, see~\cite{Aspinwall:1995mh} for details.} This is then combined with the reflection $y^i\mapsto - y^i$ on the coordinates of $T^2$ to give a freely 
acting automorphism $\gamma$ of $K3\times T^2$. The quotient of $K3\times T^2$ by this gives a Calabi-Yau manifold with Euler number zero, called the Enriques Calabi-Yau 3-fold. 
It is a K3 fibration over $\mathbb{P}^1$ with a monodromy around each of the four singularities of the base given by the Enriques involution.

The action of $\gamma$ on the charge lattice of IIA strings on $K3\times T^2$
\begin{equation}
\label{eq:622lat}
\Gamma_{6,22} \cong (E_8 \oplus U )\oplus (E_8 \oplus U) \oplus [U \oplus U \oplus U] \oplus U 
\end{equation}
is then to interchange the two $(E_8 \oplus U )$ terms, act as $-1$ on $U \oplus U \oplus U$ and to leave the final $U$ invariant.

To find the heterotic dual of the FHSV orbifold, $\Gamma_{6,22}$ is interpreted as the Narain lattice for the heterotic string compactified on $T^6$, 
with the six sub-lattices $U$ associated with the  lattice $\Gamma_{6,6}$
of heterotic momenta and winding numbers on  the six-torus. 
The action of $\gamma$ on the charge lattice and moduli space then defines  an action on the heterotic string theory (as we have done in section~\ref{sec:hetdual} 
for our models). In particular, the involution leaves one of the six circles invariant.

However the quotient of the heterotic string theory by this involution 
is not modular invariant. This was remedied in~\cite{Ferrara:1995yx} 
by supplementing the twist $\gamma$ by a shift $t$ on the  circle that is invariant under the involution. 
The shift vector $\delta $ is such that $\Delta=2 \delta \in U$ (where this $U$ is the last factor in~(\ref{eq:622lat}), 
{\it i.e.} the invariant  sub-lattice) and modular invariance requires $\Delta^2 =2$, so that 
$\delta =(1/2,1/2)$. Then   
the  shift $ y\to y+  { \pi} $  on   the circle is accompanied by a shift $ \tilde y\to \tilde y+  { \pi} $ on the dual circle. 

While this heterotic description looks quite similar to what happens in our models, in the type IIA duality frame the physics is rather different. 
The identification of the heterotic and type IIA charge lattices  under duality
relates the heterotic momentum $k$ and winding $w$ on the invariant circle with 
the type IIA D0-brane charge $x$ and the charge $z$ for D4-branes wrapping K3:
\begin{equation}
k= x, \qquad w= z
\end{equation}
In the type IIA duality frame, the action of the \lq shift' $t$  is then given as a phase rotation of the form 
\begin{equation}
\label{zphase}
\vert x,z \rangle \mapsto \exp (2\pi  i [ \alpha x  +  \beta  z ]) 
\vert x,z  \rangle = \exp (\pi  i [   x  +     z ]) 
\vert x,z  \rangle
\end{equation}
Then the IIA involution is supplemented by multiplying by the phase~(\ref{zphase}) depending on the D0-brane and D4-brane charges. 
That is, the involution $(\gamma, t)$ consists of the geometric involution on $K3\times T^2$ 
(the freely acting involution of K3 combined with the reflection on $T^2$)
supplemented by the phase rotation (\ref{zphase}).
These modifications to the Calabi-Yau compactification are visible to D-branes but not to fundamental strings, and so will not affect the perturbative type IIA string.

%\section{Moduli spaces}
%\label{sec:moduli}

%\input{moduli}

\section{\texorpdfstring{Duality Covariant Formulation and New Non-Geometric\\ Constructions}{Duality Covariant Formulation and New Non-Geometric Constructions}}
\label{sec:dualmir}

%\section{Duality Covariant Formulation and New Non-Geometric Constructions}

\subsection{Dualities and Quotients}

Suppose we have a theory $X $ on a background $M$ with a symmetry $G$, together with 
a duality map that  takes this to a theory $X' $ on a background $M'$ with a symmetry $G'$. 
Then we can consider the quotient of $X $ on $M$  by $ G $ and the quotient of $X'$ on $M'$ by $G' $ 
and ask whether they are dual, i.e. whether taking the quotient commutes with the duality transformation. 
As discussed in~\cite{Vafa:1995gm}, in general  the quotients will not be dual, but in some special cases, such as 
those in which the adiabatic argument applies, they can be dual. As usual, without a 
non-perturbative formulation of string theory the duality cannot be proved, but we can seek non-trivial tests of the duality.
     
 We have already seen here a case  where  they are not dual. Taking $X$ on $M$ to be the IIA string on 
$K3\times S^1$ and taking $G$ to be the group $\mathbb{Z}_p$ generated by a twist of the K3 CFT (corresponding 
to a mirrored automorphism) and a shift in a circle coordinate, 
then the heterotic dual of this is not modular invariant and so not consistent.
In this case we modified the heterotic symmetry $G'$ to include a winding contribution to the shift, and then made 
the dual modification to the action of $G$, involving non-perturbative NS5-brane contributions.
 Then a necessary condition for the quotients to be dual is that the group $G$ is chosen so that 
both are perturbatively consistent. Further duals could then give further non-perturbative constraints on the group $G$.

Here we are interested in  two examples: our non-geometric Calabi-Yau construction and the FHSV model for the type IIA string, 
together with the conjectured heterotic duals that were   discussed in  section~\ref{sec:mirror}. 
Consistency of the heterotic dual required modifications of the original symmetry to include D0- and D4-brane 
contributions in the FHSV model and NS5-brane contributions
 for the non-geometric Calabi-Yau construction. However, as we shall see, this is not enough to completely 
determine the non-perturbative action of the symmetry in each case.
 In our non-geometric Calabi-Yau construction, the adiabatic argument provides strong support for the duality with the heterotic T-fold.

We now turn to  the action of duality transformations on our model and that of FHSV to obtain new dual constructions.
For this, a duality covariant viewpoint is useful. 

\subsection{Compactifications to five dimensions}

We consider first compactifications to five dimensions, in both heterotic and type II duality frames. 

\subsubsection*{Symmetries and Automorphisms}

The heterotic string compactified on $T^5$ or type IIA string compactified on $K3\times S^1$ has, at generic points in the moduli space, a symmetry
\begin{equation}
[O(\Gamma _{5,21}) \ltimes U(1)^{26}]\times U(1) \, .
\end{equation}
The $U(1)^{26}\times U(1)$ is a gauge symmetry associated with $26+1$ abelian vector fields, 
and at special points in the moduli space this is enhanced to a non-abelian group.
A subgroup $U(1)^{5}$ arises from isometries of the heterotic five-torus.
The extra $U(1)$ symmetry arises in five dimensions as the NS-NS two-form  $b_2$ 
(in either the heterotic or type IIA string) can be dualised to a vector field, with a further $U(1)$ gauge symmetry that commutes with $O(\Gamma _{5,21})$. 
There are $26+1$ electric 0-brane charges  $(Z^I,K)$ corresponding to the  gauge symmetry, with $Z^I$ transforming as the 26-dimensional 
representation of $O(5,21)$.The charge $K$ is a singlet under $O({5,21})$; the 5-dimensional supersymmetry algebra has 5+1 central charges, consisting of  the 
5 electric charges for the $U(1)^5$  gauge symmetry associated with the  gauge fields in the supergravity multiplet and the singlet charge $K$. 

In the heterotic string, the BPS states carrying the charge $K$ are heterotic five-branes wrapping $T^5$.
This charge can be thought of as the winding number on $S^1$ of the solitonic string obtained from wrapping the heterotic five-brane on $T^4$.
The solitonic string of the heterotic theory is dual to the fundamental string of the type IIA theory, so in the type IIA theory the singlet 
charge $K$ is   the winding number of fundamental type IIA strings on the $S^1$ in $K3\times S^1$.
In the IIA string  theory on $K3\times S^1$ there is not a T-duality relating the winding number $K$   to the momentum on 
$S^1$, as that T-duality is not a proper symmetry of the IIA theory, but instead maps the IIA string  theory on $K3\times S^1$ 
to the IIB string  theory on $K3\times S^1$.

We are interested in automorphisms that consist of  a twist $\gamma \in O(\Gamma _{5,21})$ and a shift $t\in U(1)^{27}$ in which $t$ commutes with $\gamma$.
One possibility is to choose the shift $t$ to be generated by the singlet charge $K$, and then any $\gamma \in O(\Gamma _{5,21})$ can in principle be used.
Another is to choose  a sub-lattice $ \Gamma _{4,20} \oplus \Gamma _{1,1}\subset \Gamma _{5,21}$
so that the symmetry algebra has a subgroup
\begin{equation}
\label{ertew}
[O(\Gamma _{4,20}) \ltimes U(1)^{24}] \times [O(\Gamma _{1,1}) \ltimes U(1)^{2}] \, ,
\end{equation}
and to use a twist $\gamma \in O(\Gamma _{4,20})$ from the first factor and a shift $t\in U(1)^{2}$ from the second factor, and these indeed commute.
The automorphisms that we used in earlier sections are of this form.

\subsubsection*{The Heterotic String  Perspective}

The moduli space of heterotic strings compactified on $T^5$ is
\begin{equation}
\label{eq:modulispace5}
\mathscr{M}_\textsc{5d} \cong O(\Gamma_{5,21}) \backslash O(5,21) \slash \big( O(5) \times O(21)\big) \times  S^1 \times \mathbb{R}\, ,
\end{equation}
where the extra $S^1$ factor corresponds to a Wilson line for the gauge field dual to $b_2$ and the $\mathbb{R}$ factor is the zero mode of the  heterotic dilaton.
A $T^5$ CFT has a moduli space 
\begin{equation}
O(\Gamma_{5,5}) \backslash O(5,5) \slash \big( O(5) \times O(5)\big) 
\end{equation}
identified under the T-duality group $O(\Gamma_{5,5})$.
Then choosing a subgroup $O(5,5)\subset O(5,21) $ with corresponding  sublattice $\Gamma_{5,5}\subset \Gamma_{5,21}$ splits the heterotic 
degrees of freedom into degrees of freedom on $T^5$ described by a CFT  on $T^5$ and the remaining right-moving modes representing the gauge degrees of freedom.
This choice is not unique, and acting with the duality group $O(\Gamma_{5,21})$ will change the split into torus and gauge degrees of freedom.

For a twist  $\gamma \in O(\Gamma _{4,20})$ from the first factor in (\ref{ertew}) and a shift $t\in U(1)^{2}$ 
from the second factor in (\ref{ertew}), it is natural to choose a torus $T^4\times S^1$
so that the first factor of (\ref{ertew}) acts on  the heterotic string on $T^4$ and the second acts on the CFT on $S^1$.
Then the heterotic momentum $k$ and winding number $w$ on the  $S^1$ factor are the charges generating the $U(1)^2$ 
and transforming as a doublet under $O(1,1)$.
A shift generated by $(k,w)$ then gives a heterotic automorphism of the kind discussed in earlier sections.
It can in principle be augmented by a shift generated by the singlet charge $K$.
The 3 charges $(k,w,K)$ take values in a lattice $\Gamma _{1,1}\oplus \mathbb{Z}$ 
and transform as a $2+1$ under $O(1,1)$, with   $k$ and $w$ forming a doublet.

 The general construction could then involve a shift vector $\delta = (\alpha, \beta , \kappa)$ with three components, so that
 \begin{equation}
 \label{hetshift}
\delta \cdot \Pi = \alpha k+  \beta w +  \kappa K   \, .
\end{equation}
This would then lead to a charge-dependent phase $\exp (2\pi i \delta \cdot \Pi )$ in the automorphism.
The transformation generated by $K$ is non-perturbative and does  not affect the perturbative heterotic string.
Perturbative consistency requires that  $(\alpha, \beta)$ satisfy some modular invariant constraints, but places no constraint on $\kappa$. 
For the models considered in this article, the condition~(\ref{eq:modular_invariance_e}) is satisfied for $\alpha \beta = 1/p^2$. 
As we shall see, perturbative consistency of dual forms of the theory will impose further constraints on the shift.

Acting with $ O(\Gamma_{5,21})$ will transform $k$ and $w$ into two other linear combinations of the 
26 non-singlet charges, and in particular can lead to shifts that involve charges from the gauge sector.
This can also be thought of as changing the original choice of  split into $T^5$ degrees of freedom and 
gauge degrees of freedom to a new choice.
It will also transform the twist $\gamma$ to a conjugate twist $\tilde \gamma$.

Alternatively, we can take
 the shift $t$ to be generated by the singlet charge $K$, and take $\gamma \in O(\Gamma _{5,21})$.
 Then the shift is
 \begin{equation}
\delta \cdot \Pi =    \kappa K  
\end{equation}
for some $\kappa$.
This shift does not affect the perturbative heterotic string, so the perturbative construction is simply a quotient by $\gamma \in O(\Gamma _{5,21})$. 
In general, this will have fixed points and will result in a non-freely acting asymmetric orbifold of the heterotic string. This then 
restricts $\gamma$ to satisfy the constraints of~\cite{Narain:1986qm,Narain:1990mw} for the asymmetric orbifold to be modular invariant.

\subsubsection*{The Type IIA String  Perspective}

As we have seen, there are many ways of choosing a
split of the heterotic  degrees of freedom into degrees of freedom on $T^5$  and    gauge degrees of freedom.
For any such choice of $T^5$,  one can choose a $T^4\subset T^5$ in a number of ways, and for each choice one can 
dualize the heterotic $T^4$   to a type IIA K3 . Thus there are many ways of choosing a K3 moduli space as a subspace of 
the five-dimensional moduli space (\ref{eq:modulispace5}) -- the choices correspond to choosing an $O(\Gamma _{4,20})$ subgroup of
$O(\Gamma _{5,21})$ --
 and acting with $O(\Gamma _{5,21})$ will change this choice.
Then there is no canonical way of choosing which degrees of freedom are associated with K3 and which with $S^1$, and 
it can be changed by acting with $O(\Gamma _{5,21})$; 
%As a result, acting with $O(\Gamma _{5,21})$
it can result in different dual forms of a given compactification.

For a twist $\gamma \in O(\Gamma _{4,20})$, it is natural to choose
 a split such that the  twist $\gamma$  acts on   
K3 and the shift $t$ on $S^1$, and we now investigate this choice.
In the type IIA string, the NS-NS 2-form   is dualised to a vector field with charge $\hat z$. This is the
charge for NS5-branes wrapped on $K3\times S^1$. This can also be thought of as the
 winding charge for the solitonic string obtained from wrapping NS5-branes   on K3 and so is dual to the heterotic string winding number.
In addition, there is   a momentum $\hat k$ and a winding $\hat w$  of the type IIA string on the extra circle.
 %, and there is the charge $\hat z$ carried by the type IIA NS5-brane wrapping $K3\times S^1$.
 There are again 3 charges, and  duality relates these to heterotic charges: $k=\hat k$, $w=\hat z$ and $ K= \hat w$.
 Thus for the type IIA string, it is $(k,\hat z)$ that form a doublet under $O(1,1)$ and $\hat w$ is a singlet.
 
The general construction   involves a shift vector $\delta = (\alpha, \beta , \kappa)$ with three components, giving 
the heterotic shift (\ref{hetshift}) which is realised in the type IIA string as
 \begin{equation}
\delta \cdot \Pi =   \alpha k+  \beta \hat z +  \kappa \hat w \, .
\end{equation}
This shift leads to a charge-dependent phase $\exp (2\pi i \delta \cdot \Pi )$ in the automorphism.

\subsubsection*{The Type IIB String  Perspective}

T-duality on the $S^1$ takes the IIA string on $K3\times S^1$ to the IIB string on $K3\times S^1$.
If the IIB string has momentum $k_B$ and winding $\hat w_B$ on the $S^1$, and NS5-brane charge 
 $\hat z_B$ for NS5-branes wrapping $K3\times S^1$, these are related to the IIA string charges $k,\hat w, \hat z$ by
 \begin{equation}
k_B=\hat w, \qquad \hat w_B = k , \qquad \hat z_B=\hat z \, .
\end{equation}
Then the shift with shift vector $\delta = (\alpha, \beta , \kappa)$ acts on the type IIB string through
 \begin{equation}
\delta \cdot \Pi =   \alpha \hat w_B +  \beta \hat z _B+  \kappa k_B\, .
\end{equation}

\subsubsection*{Models}

Our original type IIA construction reviewed in sections~\ref{sec:typeII} and~\ref{sec:auty} had $\alpha \ne 0$. 
Perturbative consistency of the heterotic dual theory required $\beta \ne 0$, with $\alpha \beta = 1/p^2$.
Perturbative consistency of the type IIA  construction was achieved with no type IIA winding contributions, so this means it is 
consistent to take $\kappa =0$. Then with   $\alpha=\beta=1/p$   and $\kappa =0$ we obtain a theory which is modular invariant in both 
the perturbative heterotic and perturbative type IIA formulations.
Taking $\alpha=\beta=1/p$ but with  $\kappa \ne 0$, type IIA level-matching requires
$\kappa =0$ mod $p$, so that the shift $\kappa \hat w$ is by a lattice vector and so the corresponding phase is trivial. 
There is then no loss of generality in taking $\kappa =0$. In this case the perturbative IIB formulation is also consistent.

Acting  with $O(\Gamma _{5,21})$ will in general take the twist $\gamma$ to a conjugate transformation   that acts
not just on the K3 CFT but which acts on the full $K3\times S^1$ CFT. Note that for $p=2$ the action of the conjugate transformation on the string theory may include the world-sheet
parity-reversing transformation $\Omega$, leading to an orientifold, or $(-1)^{F_L}$.
A factor of $\Omega$ is needed whenever the conjugate transformation reverses the space-time parity.
 At the same time, the $O(\Gamma _{5,21})$ transformation  will rotate the charges 
$k, \hat z$ to other charges for the $U(1)^{26}$ symmetry. The singlet charge $K=  \hat w$ does not change.
(This can instead be viewed as changing which subsector of the theory is to be interpreted as corresponding to the K3 CFT.)
For example, there is a transformation that takes $k$ to the D0-brane charge $Z_0$  and 
$\hat z$ to the charge $Z_4$ for D4-branes wrapping K3.
This would give a shift 
\begin{equation}
\delta \cdot \Pi =   \alpha Z_0+  \beta Z_4 +  \kappa \hat w
\end{equation}
which is completely non-perturbative, giving a phase  rotation to any given state depending on its 
D0,D4 and NS5 charges. For the perturbative theory, this is simply a $\mathbb{Z}_p$ orbifold of the 
type IIA string on $K3\times T^2$ by $\tilde \gamma$, with $\tilde \gamma$ now acting non-trivially on 
$K3\times T^2$ (i.e. not just acting on K3).
Perturbative consistency of this then does not depend at all on 
the parameters $ \alpha , \beta ,  \kappa $ and only depends on the choice of twist $\gamma$.
However, this is still dual to the heterotic construction, and perturbative consistency of the heterotic dual constrains 
$ \alpha $ and $ \beta$, as above. Similarly, the original IIA version sets $\kappa=0$.

Finally, we can instead  take
 the shift $t$ to be generated by the singlet charge $K$, and take $\gamma \in O(\Gamma _{5,21})$.
 Then the shift becomes
 \begin{equation}
\delta \cdot \Pi =    \kappa  \hat w  
\end{equation}
for the type IIA string. $ \hat w  $ is a perturbative charge for the type IIA string, but it is not 
constrained by IIA modular invariance since the shift vector involves a winding charge but no momentum. In this 
case, the only constraint is that $p \kappa  \hat w  
$ is a lattice vector, so $\kappa = n/p$ for some integer $n<p$.

\subsection{Compactifications to four dimensions}

We now turn to compactifications to four dimensions, which allow more general constructions. 

\subsubsection*{Symmetries and Automorphisms}

The heterotic string compactified on $T^6$ or type IIA string compactified on $K3\times T^2$ has, at generic points in the moduli space, a symmetry
\begin{equation}
\label{4dsymm}
[O(\Gamma _{6,22})\times SL(2, \mathbb{Z})] \ltimes U(1)^{56} \, .
\end{equation}
There is a $U(1)^{28}
$  gauge symmetry associated with   $28$ gauge fields, 
and, formally,  a further $U(1)^{28}
$   symmetry associated with the
S-dual gauge fields. In different S-duality frames,  different  subgroups $U(1)^{28}\subset U(1)^{56}$ will be realised as  fundamental  gauge symmetries.
There are 28 electric and 28 magnetic charges, transforming in the $(28,2)$  representation under $O(6,22)\times SL(2)$.

Here we will focus on 
twists in $O(\Gamma _{6,22}) $ 
and not consider twists involving S-duality.
The discussion is then very similar to the 5-dimensional case above.
%At special points in the moduli space, $U(1)^{28}$ is enhanced to a non-abelian group.
%a subgroup $U(1)^{6}$ arises from isometries of the heterotic six-torus.
We will consider an automorphism $(\gamma,t)$  consisting of a twist $\gamma \in O(\Gamma _{6,22})$ and a shift $t\in U(1)^{56}$ where 
$t$ commutes with $\gamma$.

Choosing a sub-lattice $\Gamma _{5,21} \oplus \Gamma _{1,1}\subset \Gamma _{6,22}$, 
the symmetry algebra of the theory has a subgroup
\begin{equation}
\label{512split}
[O(\Gamma _{5,21}) \ltimes U(1)^{52}] \times [O(\Gamma _{1,1}) \ltimes U(1)^{4}]\, .
\end{equation}
We can then use a twist $\gamma \in O(\Gamma _{5,21})$ from the first factor and a shift $t\in U(1)^{4}$ from the second factor, and these indeed commute.

We will also consider choosing a sub-lattice
$ \Gamma _{4,20} \oplus \Gamma _{2,2}\subset \Gamma _{6,22}$, 
selecting   a subgroup of the symmetry algebra 
\begin{equation}
\label{420split}
[O(\Gamma _{4,20}) \ltimes U(1)^{48}] \times [O(\Gamma _{2,2}) \ltimes U(1)^{8}]
\end{equation}
 and using a twist $\gamma \in O(\Gamma _{4,20})$ from the first factor and a shift $t\in U(1)^{8}$ from the second factor.
 One class of examples arises in taking a reduction to 5 dimensions of the kind considered in the previous subsection, with a twist $\gamma \in O(\Gamma _{4,20})$ and a shift on a circle, followed by a standard reduction (no twist or shift) on a further circle;
 such cases have been the main focus  in this paper.
 We can also consider reductions by 
 $(\gamma_1,t_1)$ and $(\gamma_2,t_2)$
 where $\gamma_1, \gamma_2$ are two commuting twists in $O(\Gamma _{4,20})$ and $t_1,t_2$ are two shifts in $U(1)^{8}$  (see~\cite{Hull:2017llx} for an 
analysis of models with two twists). Note that the 8-charges for $U(1)^{8}$
 transform as a
 $(4,2)$ under $O(2,2)\times SL(2)$. Using $O(2,2) \sim  SL(2)\times SL(2)$, this is the
 $(2,2,2)$ representation of $SL(2)\times SL(2) \times SL(2)$.

  \subsubsection*{The Heterotic String  Perspective}
  
 Consider first the case with a twist $\gamma \in O(\Gamma _{5,21})$ from the first factor in  (\ref{512split})  and a shift $t\in U(1)^{4}$ from the second factor in  (\ref{512split}).
 It is natural to choose a split so that  the sub-lattice $ \Gamma _{5,21}$ is associated with  the heterotic string compactified on $T^5$ and the sub-lattice 
$\Gamma _{1,1} $ with a further circle compactification. The charges for the
$U(1)^{4}$ symmetry are the heterotic momentum $k$ and winding $w$ on the extra circle, the heterotic 5-brane charge $z$ for heterotic 5-branes wrapping $T^5$ and the Kaluza-Klein (KK) monopole charge 
$q$.\footnote{The KK monopole charge arises from   solutions of the form
$\mathbb{R} \times ALF \times T^5$ where $\mathbb{R}$ is a timelike direction and $ALF$ denotes an ALF gravitational instanton with charge $q$ (so that for $q=1$ we have self-dual Taub-NUT space). The   \lq extra circle' is the fibre of the ALF gravitational instanton.}
 
Then the general shift vector is given by $\delta = (\alpha, \beta , \lambda, \kappa)$ with four components, so that
 \begin{equation}
 \label{hetshifta}
\delta \cdot \Pi = \alpha k+  \beta w + \lambda q+ \kappa z  \, .
\end{equation}
The shifts involving $z,q$ do not affect the perturbative theory. For the models of section~\ref{sec:hetdual} --~with $\gamma \in O(\Gamma _{4,20})$~--
perturbative consistency is  achieved if both $\alpha $ and $\beta $ are non-zero with $\alpha \beta = 1/p^2$. A similar analysis can be done for the 
general case with arbitrary $\gamma \in O(\Gamma _{5,21})$. 
%In the particular case in which $\gamma \in O(\Gamma _{4,20})$, then the constraints from section 4 are sufficient for modular invariance; 
%similar results hold for the general case.

The case with a twist $\gamma \in O(\Gamma _{4,20})$ from the first factor in  (\ref{420split})  
and a shift $t\in U(1)^{8}$ from the second factor in  (\ref{420split}) is very similar.
Choosing the natural split in which the sub-lattice $ \Gamma _{4,20}$ is associated with 
the heterotic string compactified on $T^4$ and the sub-lattice $\Gamma _{2,2} $ with a further $T^2$ compactification, the
charges for the
$U(1)^{8}$ symmetry are the heterotic momenta $k_i$ and windings $w^i$ on the $T^2$,
 the heterotic 5-brane charges $z_i$ for heterotic 5-branes wrapping $T^5$ and the Kaluza-Klein monopole charges 
$q^i$, 
where $i=1,2$ is a   coordinate index on $T^2$, which has coordinates $y^i$.
The charge $z_i$ is for a  5-brane wrapping the $y^i$ circle and the $T^4$, 
so it is the winding number for the solitonic string from the 5-brane wrapping $T^4$.
The general shift is then of the form
 \begin{equation}
 \label{hetshiftib}
\delta \cdot \Pi = \alpha ^i k_i+  \beta_i w^i + \lambda _iq^i+ \kappa ^iz_i  
\end{equation}
For the models of section~\ref{sec:hetdual}, perturbative consistency requires $\alpha ^i    \beta_i =1/p^2$.
Taking the only non-zero coefficients to have, say, $i=1$ reduces this to the previous case.

 \subsubsection*{The Type IIA String  Perspective}

For the case with a twist $\gamma \in O(\Gamma _{5,21})$ from the first factor in~(\ref{512split})  
and a shift $t\in U(1)^{4}$ from the second factor in~(\ref{512split}),
the natural   choice of split has the lattice $ \Gamma _{5,21}$  associated with  the type IIA  
string compactified on $K3\times S^1$ and the lattice $\Gamma_{1,1} $ associated with a further 
compactification on a circle with coordinate $y^1$. In this case the $U(1)^4$ charges  are the momentum $\hat k$ and type IIA winding $\hat w$ on the  $y^1$ circle, 
the charge $\hat z$ from an 
NS5-brane wrapping K3 and the $y^1$ circle, and the 
KK monopole charge $\hat q$ associated with the $y^1$ circle.

Then heterotic-type II duality relates these to heterotic charges: $k=\hat k$, $w=\hat z$ and $ z= \hat w$, 
$q=\hat q$.
The general shift vector  $\delta = (\alpha, \beta , \lambda, \kappa)$  gives (\ref{hetshiftib}) in the heterotic picture and 
  \begin{equation}
\delta \cdot \Pi =   \alpha \hat k+  \beta \hat z +\kappa    \hat w + \lambda  \hat q
\end{equation}
for type IIA.
This shift again leads to a charge-dependent phase $\exp (2\pi i \delta \cdot \Pi )$ in the automorphism.
Level matching of the perturbative type IIA string with
$\alpha \ne 0$ 
leads to $\kappa=0$, as in the five-dimensional analysis above but places no constraints on
$\beta, \lambda$ as they correspond to non-perturbative contributions for the IIA string.
Requiring perturbative consistency of both the IIA and heterotic formulations is satisfied (for the models of section~\ref{sec:hetdual}) with 
$\alpha=\beta=1/p$,   $\kappa= 0$ but puts no constraints on $\lambda$.
The perturbative IIB formulation gives no further constraints. 

As in the five dimensional case, we can consider acting on a dual pair with a duality transformation.
This will transform the charges appearing in the shift, and take the twist to a conjugate one, which for $p=2$ might include factors of $\Omega$ or $(-1)^{F_L}$.

\subsubsection*{  S-Duality}

To find a constraint on the parameter $\lambda$, one could seek a duality that transforms $q$ to a perturbative charge that would enter 
into the perturbative constraints in the dual theory. Such a duality is provided by the heterotic string S-duality.

The heterotic charges $(k,w,z,q)$ transform as a $(2,2)$ under 
$O(\Gamma _{1,1})\times SL(2,\mathbb{Z})$, with $(k,w)$ and $(z,q)$ transforming as doublets under the T-duality $O(\Gamma _{1,1})$
and $(k,q)$ and $(w,z)$ transforming as doublets under the S-duality $SL(2,\mathbb{Z})$.
Then  acting with the  $SL(2,\mathbb{Z})$ element
$$\begin{pmatrix}
		0 & 1 \\
		-1 & 0 
	\end{pmatrix}$$
takes the shift (\ref{hetshifta}) to
 \begin{equation}
 \label{hetshiftsd}
\delta \cdot \Pi = -\kappa  w-  \lambda k +
\alpha q+  \beta z  \, .
\end{equation}
while leaving the twist unchanged.
If we were to demand
perturbative consistency of this S-dual theory, this would be  achieved only if both $\lambda $ and $\kappa $ are non-zero with $\lambda \kappa = 1/p^2$ once again for the models of section~\ref{sec:hetdual}.
We then learn that perturbative consistency of the heterotic string and of the S-dual
heterotic string would require all four components of the shift vector to be non-zero, and we could satisfy these requirements by  taking 
\begin{equation}
\alpha=\beta = \lambda =\kappa = \frac 1 p
\end{equation}

However, in this case S-duality doesn't commute with the quotient -- the strong coupling behaviour of the $\mathcal{N}=2$ supersymmetric theory arising from the quotient is not given by the strong coupling behaviour of the original  $\mathcal{N}=4$ supersymmetric theory. Then the constraint $\lambda \kappa = 1/p^2$ should not be applied to the original theory, and we can keep $\kappa=0$, as found above.

One can see directly why the adiabatic argument fails in this case. Heterotic S-duality corresponds, in type IIA variables, to a double T-duality on the two-torus, sending 
the torus area $A$ to $(\alpha')^2/A$.  The adiabatic argument holds in the limit where the $T^2$ base is large, hence is not compatible with this duality transformation.

\subsubsection*{The  FHSV Model Revisited}

The lattice
$\Gamma_{5,21}$ is given by
\begin{equation}
\Gamma_{5,21} \cong E_8 \oplus E_8 \oplus U \oplus U \oplus U \oplus U \oplus U\, .
\end{equation}
Consider then the automorphism $\gamma$ given by
interchanging two $E_8 \oplus U$ sublattices and acting as $-1$ on the remaining  
sublattice $U\oplus U \oplus U$. This twist
  $\gamma \in O(\Gamma _{5,21})$ can be associated with  the first factor in~(\ref{512split}) 
and combined with a shift $t\in U(1)^{4}$ from the second factor in  (\ref{512split}), with
shift vector  $\delta = (\alpha, \beta , \lambda, \kappa)$.
  
The heterotic string dual of the FHSV model discussed in subsection~\ref{subsec:fhsv} is of precisely this form. With the natural 
choice of split in which the  the sub-lattice $ \Gamma _{5,21}$ is associated with  the 
heterotic string compactified on $T^5$ and the sub-lattice $\Gamma _{1,1} $ with a further circle compactification, the shift is 
$\delta \cdot \Pi = \alpha k+  \beta w + \lambda q+ \kappa z $. Perturbative consistency required both $\alpha, \beta $ 
to be non-zero with $\alpha \beta = 1/4$~\cite{Ferrara:1995yx}.

In the FHSV model, $\gamma$ is not taken to act on $K3\times T^2$ in the way we have referred to as \lq natural'.
In choosing the sub-lattice $ \Gamma _{5,21} \oplus \Gamma _{1,1}\subset \Gamma _{6,22}$, 
we take the $\Gamma _{1,1}$ part of the charge lattice to be the one corresponding to D0-brane charge 
and D4-brane charge (for D4-branes wrapping K3). 
Then $\gamma$ acts on K3 through the Enriques involution and on $T^2$ as a reflection. 
For the model of~\cite{Ferrara:1995yx,Aspinwall:1995mh}, 
the shift was taken to be
 \begin{equation}
\delta \cdot \Pi =   \alpha Z_0+  \beta Z_4 \, ,
\end{equation}
where $Z_0$ is the D0-brane charge and $Z_4$ the charges of D4-branes wrapping K3, giving 
a phase  rotation to any given state depending on its D0 and D4 charges. 
The general 
heterotic shift $\delta \cdot \Pi = \alpha k+  \beta w + \lambda q+ \kappa z $
 would correspond to extending
the FHSV construction must be extended by taking $ \lambda, \kappa$ non-zero,
giving
\begin{equation}
\delta \cdot \Pi =   \alpha Z_0+  \beta Z_4  +
\kappa  Z_2 +  \lambda Z_6
\end{equation}
where $Z_2$ is the charge for D2-branes wrapping $T^2$ and $Z_6$ is the charge for D6-branes wrapping $K3\times T^2$.
Perturbative consistency of the FHSV construction places no constraint on the four parameters.

However, we can instead make the following choice, giving a type IIA  dual of the heterotic FHSV model which looks 
different from the original Enriques Calabi-Yau type IIA compactification. In choosing the sub-lattice 
$ \Gamma _{5,21} \oplus \Gamma _{1,1}\subset \Gamma _{6,22}$, 
we now take $ \Gamma _{5,21}$ to be the charge lattice for the IIA string on $K3\times S^1$, so that
  $\gamma$   acts as an involution of $K3\times S^1$, with a fixed point locus, and 
$\Gamma _{1,1}$ is associated with a further circle reduction. Then the shift is 
\begin{equation}
\label{orientishift}
\delta \cdot \Pi =   \alpha \hat k+  \beta   \hat z +  \kappa  \hat w+   \lambda \hat q
\end{equation}
where  $\hat k$ is the IIA momentum,  $ \hat z$ is
the NS5-brane charge.
$\hat w$ is the IIA string winding number and $ \hat q
$ is the KK monopole charge.
The perturbative charges are $\hat k, \hat w$. 

In this case, the transformation given by this twist and shift is not quite a symmetry of the IIA string on $K3\times T^2$.
The twist involves a reflection $y\to -y$ on the circle in $K3\times S^1$ and this 
must be combined with a world-sheet parity transformation $\Omega$ to give a symmetry.
We then have an orientifold of the  IIA string on $K3\times T^2$ by $\Omega$
combined with the shift and  twist described above. For the shift
$\delta \cdot \Pi =   \alpha \hat k$ this is  an  orientifold analysed in~\cite{Vafa:1995gm}
as a dual to the FHSV model. 
We can consider generalising  this by extending the  shift to (\ref{orientishift}). 
Perturbative consistency of the IIA theory then leads to $ \kappa=0$ as before. The heterotic dual gives $\alpha \beta = 1/4$.
As noted in~\cite{Vafa:1995gm}, the adiabatic argument supports the duality between this orientifold and the heterotic dual, but does not apply to the duality between the FHSV model and its heterotic version.

This type IIA orientifold is non-geometric, following the analysis of section~\ref{sec:typeII}; 
the action on the K3 CFT is in $O(\Gamma_{4,20})$ but not in $O(\Gamma_{3,19})$, and the shift corresponding to the second circle 
has both momentum and winding components. However through heterotic/type IIA duality it is expected to be non-perturbatively equivalent 
to type IIA compactified on the Enriques Calabi-Yau threefold.

\subsection{Non-Geometric Constructions}

The general class of construction we have been discussing consists of a quotient of a string theory background by a twist $\gamma$ of order $p$ in a duality group $O(\Gamma_{n,n+16})$ for $n=4$ or $n=5$ together with a shift $t$.
From the discussion in section 2, when $t$ is a simple shift $t:y\mapsto y+ 2\pi/p$ of a circle coordinate $y$, this can be seen as a special point in the moduli space of a duality twisted reduction, with the dependence of all fields on $y$ given by a continuous duality transformation
 $g(y)\in  O(n,n+16)$ with monodromy $\gamma$. If the monodromy transformation acts geometrically, this constructs a bundle over a circle with fibre $T^4$ or $T^5$ or K3 or $K3\times S^1$. For example, starting from type IIA compactified on $K3\times S^1$ with $\gamma \in O(\Gamma_{3,19})\subset O(\Gamma_{4,20})$ acting as a K3 diffeomorphism, the duality twisted reduction can be understood as a geometric compactification of the type IIA string
on a K3 bundle over $S^1$. More generally, the result is non-geometric. If $\gamma$  involves  T-duality transformations, we have a T-fold and if it includes  mirror transformations, we have a mirror-fold.

Our heterotic construction involved a shift vector $\delta = (\alpha, \beta )$
so that the shift is generated by
 \begin{equation}
 \label{hetshiftac}
2 \pi i \delta \cdot \Pi = \alpha k+  \beta w  \end{equation}
with $\alpha \beta = 1/p^2$.
As we have seen in section~\ref{sec:mirror}, this can be thought of as acting as  a phase rotation  on a state with  
momentum $k$ and heterotic winding number $w$, or as giving a shift on the 2-dimensional doubled circle with coordinates $y, \tilde y$ with  $y\to y+ 2\pi \alpha$, $\tilde y \to  \tilde y +2\pi \beta$.
The theory can be formulated as a double field theory with fields depending on both $y$ and $ \tilde y$. 
Then this is a special point in the moduli space of a duality twisted reduction 
in which the dependence of all fields on $y, \tilde y$ is given by a continuous duality transformation
 $g(y, \tilde y)\in  O(n,n+16)$ with monodromy $\gamma$:
 \begin{equation}
g(y, \tilde y)^{-1} g(y+ 2\pi \alpha, \tilde y+2\pi \beta)=\gamma
\end{equation}
(This is a special case of a more general construction in which there could be different monodromies in the $y$ and $ \tilde y$ directions.)
For geometric monodromy in $GL(n, \mathbb{Z})$ acting as a diffeomorphism of $T^n$,  this constructs a $T^n$ bundle over the doubled circle, while for   a T-duality monodromy in $O(\Gamma_{n,n})$ this constructs a bundle of a 2n-dimensional doubled $n$-torus over the doubled circle, which gives the geometric realisation of a T-fold in the doubled formalism~\cite{Hull:2004in}.  For a general monodromy in $O(\Gamma_{n,n+16})$, this is a bundle with fibre the heterotic doubled torus $T^{n,n+16}$ of dimension $2n+16$ over the 2-dimensional doubled circle, which can be regarded as a configuration for heterotic double field theory.

Our general construction involved further charges $Q_I$, so that the shift vector was of the form $\delta = (\alpha, \beta , \lambda ^I )$
 \begin{equation}
 \label{genshift}
2 \pi i \delta \cdot \Pi = \alpha k+  \beta w + \lambda ^I Q_I
 \end{equation}
 These too can be geometrised by going to an extended field theory with further coordinates
 $u^I$ on which the charges $Q_I$ act as translations:
 \begin{equation}
Q_I= -i \frac {\partial } {\partial u^I}
\end{equation}
Then in the extended field theory, the coordinates that the fields depend on include $y, \tilde y, u^I$ and the shift $t$ acts as a translation on $y, \tilde y, u^I$, resulting in a generalised bundle over a base space (typically a torus) with coordinates $y, \tilde y, u^I$.

\section{Conclusion}
\label{sec:concl}

In this paper, we have proposed a   four-dimensional $\mathcal{N}=2$ non-perturbative duality
relating non-geometric Calabi-Yau compactifications of the type IIA superstring to T-fold compactifications of the heterotic superstring and have shown that this duality follows from the adiabatic argument.
The non-geometric type II backgrounds were constructed in~\cite{Hull:2017llx} as $K3$ fibrations over $T^2$ with monodromy twists associated with the action of 
{\it mirrored K3 automorphisms} on the $K3$ CFT. 
The K3 automorphisms are realised in the heterotic string as element of the T-duality group, 
and the heterotic duals are $T^4$ fibrations over $T^2$ with T-duality monodromy twists.
At points in the moduli space which are fixed under the action of the monodromy automorphisms, 
the construction reduces to an asymmetric orbifold on the heterotic side and to an 
asymmetric Gepner model in type IIA.
At these fixed points, there is no enhanced gauge symmetry but there is enhanced discrete symmetry.

These models preserve $\mathcal{N}=2$ supersymmetry in four dimensions. The automorphism acts on the lattice $\Gamma_{4,20}$ 
by an isometry in $O(\Gamma_{4,20} )$ that leaves no sublattice of $\Gamma_{4,20} $ invariant.  
For the heterotic string on $T^4$, all of the four left-moving and twenty right-moving  chiral bosons transform. 
On the type II side of the duality, the D-brane charge lattice is $\Gamma_{4,20}$ and the fact 
that no sublattice is left invariant by the twist means that all BPS D-brane states are projected out by the orbifold. 
This is consistent with the fact that there are no Ramond-Ramond ground states in these theories, since all space-time
supersymmetry comes from the left-movers.

The naive heterotic dual of the type IIA construction is not modular invariant. We found a modification of the 
heterotic construction that is modular invariant, and this modification led in turn to a 
non-perturbative modification of the type IIA model. A similar story applies to the FHSV model.
 For the type IIA string, the modification can be viewed as necessary for non-perturbative consistency. Although we do not have a complete 
non-perturbative formulation, it seems that a necessary condition for the  non-perturbative consistency of a model should be that the theory 
is modular invariant in all possible duality frames, and in any given frame  this can require non-perturbative corrections, as we have seen. 
Our models are perturbatively consistent in the IIA, IIB and heterotic duality frames.
Acting with a duality transformation then takes us to a new perturbative theory (which can also be thought of as choosing a different 
modulus of the original theory as a coupling constant) and we again require consistency in this new perturbative theory.

Let us explore this further.
It is believed there is a non-perturbatively consistent string solution that can be treated as a perturbation theory in terms of the IIA 
coupling constant, the IIB coupling constant or the heterotic  coupling constant. The perturbative heterotic theory is the heterotic 
string compactified on $T^6$, while the perturbative IIA (IIB) theory is the IIA (IIB) string compactified on $K3\times T^2$.
The theory is believed to have an exact non-perturbative symmetry~(\ref{4dsymm}) and we are interested in taking quotients of the 
theory by a $\mathbb{Z}_p$ subgroup of this. The key question is which $\mathbb{Z}_p$ subgroups  lead to consistent theories. 
We have seen that different restrictions arise from requiring perturbative consistency as a IIA, IIB or heterotic theory. 
Acting with a symmetry in~(\ref{4dsymm}) maps the  $\mathbb{Z}_p$ subgroup to a conjugate $\mathbb{Z}_p$ subgroup embedded 
differently in the symmetry group and gives a new quotient. The new quotient will not in general be dual to 
the original one, but in an important class of cases, such as the ones studied here in which the adiabatic argument can be applied,  
this gives a new dual  of the original construction.
 Perturbative consistency of each such dual theory gives further constraints. 
 In this way, we find a set of necessary conditions for the consistency of the quotient.
 Knowing whether these are sufficient would require an understanding of the non-perturbative 
theory, but these conditions give us  important information about the non-perturbative theory that it would be interesting to investigate further.

The $\mathbb{Z}_p$ symmetries we have been quotienting by are generated by a transformation $(t,\gamma)$ consisting of a twist
$\gamma \in O(\Gamma _{5,21})$   and a shift $t\in U(1)^{4}$ (or a  twist
$\gamma \in O(\Gamma _{4,20})$   and a shift $t\in U(1)^{8}$). The adiabatic argument led us to use the same twist in each duality frame, but we 
found different consistency conditions on the shift in different  duality frames. In our original IIA construction, the shift was a simple 
order-$p$ shift of a circle coordinate $y\mapsto y+ 2\pi/p$ and this was sufficient for IIA modular invariance.
For the heterotic dual, heterotic modular invariance required also shifting the T-dual coordinate $\tilde y\mapsto \tilde y+ 2\pi/p$, or, equivalently, the action of $t$ on a state with momentum $k$ and winding $w$ on the circle was to multiply by a phase $\exp (2\pi i (k+w)/p)$. Transforming back to the IIA theory, the heterotic winding number $w$ is mapped to the NS5-brane wrapping number, and so action of $t$ on the IIA string involves a phase depending on the NS5-brane charge, giving a non-perturbative modification of the theory.
The general picture involves a phase depending on four charges  for a shift $t\in U(1)^{4}$ or eight charges for a shift $t\in U(1)^{8}$, and acting with a duality transformation can change which charges they are. For example, the FHSV construction involved a phase depending on the D0- and D4-brane charges, while the dual we found had a phase depending on the type IIA momentum and NS5-brane charge.

We now return  to Harvey and Moore's  question. 
There are two classes of $\mathcal{N}=2$ heterotic toroidal orbifolds with a 
known type II dual: 
quotients by symmetries that preserve a D-brane charge lattice,  corresponding to IIA on Calabi-Yau three-folds, and quotients that do not preserve any charge 
lattice,  corresponding to non-geometric compactifications based on mirrored automorphisms. 
In general, an orbifold of the heterotic string on $T^{4+n}$ by a symmetry $G$ is mapped to an orbifold of the type IIA string on $K3\times T^n$ by the dual  $ G$  which  now acts on the type IIA string; this will require that the orbifold is non-perturbatively consistent, so that in particular it is modular invariant in both the heterotic and type IIA duality frames.
Consider  for example a general $\mathbb{Z}_p$  orbifold of the heterotic string on $T^4\times S^1$ by $(\gamma,t)$, where $\gamma \in O(\Gamma _{4,20})$ acts as a heterotic T-duality and the shift gives a phase depending on the momentum and the heterotic winding number on the $S^1$.
This is then mapped to 
an orbifold of  the type IIA string on $K3\times S^1$ by the
transformation  $(\gamma,t)$ in which $\gamma$ acts as a K3 automorphism and $t$ gives a phase depending on the momentum on the $S^1$ and the NS5-brane charge
for NS5-branes wrapping $K3\times S^1$.
In some cases the type IIA dual is a CY compactification, but in general it will lead to a non-geometric construction.
It will be interesting to explore this duality further, for example for the models of~\cite{Israel:2013wwa,Israel:2015efa,Blumenhagen:2016axv}.

\subsection*{Acknowledgments}

We thank Miranda Cheng, Alessandra Sarti, Katrin Wendland and especially Boris Pioline for discussions and correspondence. 
The work of Y.G., C.H. and D.I. received support from the ILP LABEX (ANR-10-LABX-63) 
supported by French state funds managed by the ANR (ANR-11-IDEX-0004-02) and from the project QHNS in the program ANR Blanc 
SIMI5 of Agence National de la Recherche. 
The  work of CH was supported  by  the EPSRC programme grant 
``New Geometric Structures from String Theory" EP/K034456/1 and the STFC  grant ST/L00044X/1.

\appendix

\section{Partition Function Computations}
\label{app:part_function}

\subsection*{\texorpdfstring{$\vartheta$}{Theta} functions}

In this section, we give our conventions for $\vartheta$ functions and recall some of their modular properties that are  useful in our computations. We define the Jacobi $\vartheta$ function with characteristic as

\begin{equation}
  \vartheta\jtheta{\alpha}{\beta} \left(\tau \middle| v \right) := \sum_{n \in \mathbb{Z}} q^{\frac{1}{2} \left(n+\frac{\alpha}{2}\right)^2} e^{2 i \pi \left(n + \frac{\alpha}{2}\right)\left(v+\frac{\beta}{2}\right)} ,
\end{equation}
where $\alpha, \beta \in \mathbb{R}$ and where   $q$ is defined, as usual, by $q := \exp(2 i \pi \tau)$.   $\vartheta$ also admits the product representation~\cite{Blumenhagen:2013fgp}

\begin{equation}
  \frac{\vartheta\jtheta{\alpha}{\beta}(\tau|v)}{\eta(\tau)} = e^{i \pi \alpha\left(v+\frac{\beta}{2}\right)} q^{\frac{\alpha^2}{8}-\frac{1}{24}} \prod_{n=1}^\infty \left(1+q^{n+\frac{\alpha-1}{2}} e^{2 i \pi \left(v+\frac{\beta}{2}\right)}\right) \left(1+q^{n-\frac{\alpha+1}{2}} e^{-2 i \pi \left(v+\frac{\beta}{2}\right)}\right) ,
\end{equation}
where $\eta(\tau)$ is the Dedekind $\eta$ function defined by

\begin{equation}
  \eta(\tau) := q^{1/24} \prod_{n=1}^\infty (1-q^n).
\end{equation}
The well-known modular properties of the
$\vartheta$ functions   makes them functions are  a powerful tool in constructing modular invariant quantities. Their behaviour under the generators of $SL(2, \mathbb{Z})$ are given by

\begin{subequations}
  \begin{equation}
    \vartheta\jtheta{\alpha}{\beta} \left(\tau + 1 \middle| v \right) = e^{-\frac{i\pi}{4} \alpha(\alpha-2)} \vartheta\jtheta{\alpha}{\alpha + \beta - 1} \left(\tau \middle| v \right) ,
  \end{equation}
  \begin{equation}
    \vartheta\jtheta{\alpha}{\beta} \left(-\frac{1}{\tau} \middle| \frac{v}{\tau} \right) = e^{\frac{i\pi}{2} \alpha \beta + \frac{i\pi}{\tau} v^2} \vartheta\jtheta{-\beta}{\alpha} \left(\tau \middle| v \right).
  \end{equation}
\end{subequations}
It is also easy to show that the arguments $\alpha$ and $\beta$ satisfy the periodicity properties

\begin{equation}
  \vartheta\jtheta{\alpha +2}{\beta} \left(\tau \middle| v \right) = \vartheta\jtheta{\alpha}{\beta} \left(\tau \middle| v \right) \quad \mathrm{and} \quad \vartheta\jtheta{\alpha}{\beta + 2} \left(\tau \middle| v \right) = e^{i \pi \alpha} \vartheta\jtheta{\alpha}{\beta} \left(\tau \middle| v \right).
\end{equation}
In the following, we will drop the explicit $\tau$ dependence of the $\vartheta$ functions and write only $\vartheta\jtheta{\alpha}{\beta}(v)$ (or simply $\vartheta\jtheta{\alpha}{\beta}$ if $v=0$). An especially useful identity when it comes to computing BPS indices for instance is the famous Jacobi abstruse identity which allows one to sum over spin structures and which reads~\cite{Kiritsis:1997hj}

\begin{equation}
  \frac{1}{2} \sum_{\alpha,\beta=0}^1 (-1)^{\alpha+\beta+\alpha\beta} \prod_{i=1}^4 \vartheta\jtheta{\alpha+h_i}{\beta+g_i}(v_i) = -\prod_{i=1}^4 \vartheta\jtheta{1-h_i}{1-g_i} (v_i')
\end{equation}
provided that $\sum_i h_i = \sum_i g_i = 0$; here,

\begin{equation}
  \begin{aligned}
    v'_1 = \frac{1}{2}(-v_1 + v_2 + v_3 + v_4) \quad , \quad v'_2 = \frac{1}{2}(v_1 - v_2 + v_3 + v_4) \\
    v'_3 = \frac{1}{2}(v_1 + v_2 - v_3 + v_4) \quad , \quad v'_4 = \frac{1}{2}(v_1 + v_2 + v_3 - v_4).
  \end{aligned}
\end{equation}

\subsection*{Restrictions on the shift vector}

Let us first show that it is always possible to find a representative of the shift vector in $\frac{1}{p} \Gamma_{2,2} \slash \Gamma_{2,2}$ such that 
\begin{equation}
    \label{eq:modular_invariance_es}
 p^2 \alpha ^i \beta _i = \Psi_p\, ,
\end{equation}
so
equation~\eqref{eq:modular_invariance_e} holds strictly,   not just modulo $p$. 
First, one may note that $\Psi_p$ and $p$ must be coprime, as follows from $\gcd(s,p)=1$ - the latter being imposed by equation~\eqref{eq:automorphism_eigenvalues}. Assuming that one starts with a shift vector $\delta$ satisfying equation~\eqref{eq:modular_invariance_e}, this means now that $\gcd(\alpha^1,\alpha^2,p)=1$ as well since equation~\eqref{eq:modular_invariance_e} would not admit a solution for $\beta_1$ and $\beta_2$ otherwise. In such a case, it is always possible to define $\widetilde{\alpha}^1 := \alpha^1$ and $\widetilde{\alpha}^2 := \alpha^2 + t p$ for some integer $t$ so that $\gcd(\widetilde{\alpha}^1,\widetilde{\alpha}^2)=1$; indeed, the existence of a solution to $$\left\lbrace\begin{array}{l l} t = 1\ \mathrm{mod}\ q \quad & \forall\ \mathrm{prime}\  q\ | \gcd(\alpha^1,\alpha^2) \\ t = 0\ \mathrm{mod}\ q' & \forall\ \mathrm{prime}\ q'\ |\ \alpha^1\ \mathrm{and}\ q' \nmid \gcd(\alpha^1,\alpha^2)\end{array}\right.$$ is guaranteed by the Chinese remainder theorem\footnote{Which proves more generally the existence of a solution to $$x = x_i \ \mathrm{mod}\ p_i \quad i=1,...,n$$ for any set or pairwise coprime integers $\lbrace p_i , i=1,...,n\rbrace$.}, and one may show that such an integer $t$ would lead to $\gcd(\widetilde{\alpha}^1,\widetilde{\alpha}^2)=1$ as required. B\'ezout's identity\footnote{Which states that $$\alpha x + \beta y = \gamma$$ admits a solution for $(\alpha,\beta)$ if and only if $\gcd(x,y)\, |\, \gamma$; in particular, it therefore ensures the existence of solutions to the above equation for any integer $\gamma$ in the case where $x$ and $y$ are coprime integers.} then finally ensures us that there exist integers $\widetilde{\beta}_1$ and $\widetilde{\beta}_2$ with $\widetilde{\beta}_i = \beta_i\ \mathrm{mod}\ p$ such that $\widetilde{\alpha}^i \widetilde{\beta}_i = \Psi_p$, so that we can indeed choose a representative of any given vector shift $\delta$ satisfying~\eqref{eq:modular_invariance_e} strictly.

We now   give more details about how one gets to equation~\eqref{eq:modular_invariance_c}. First, it may be shown (see \textit{e.g.}~\cite{apostol_introduction_1976}) that $$\sum_{\substack{a = 1 \\ \gcd(a,p)=1}}^p a^k = \sum_{d | p} \mu\left(\frac{p}{d}\right) \left(\frac{p}{d}\right)^k \sum_{a=1}^d a^k$$ for any integers $k$ and $p$, $\mu$ being here the M\"obius function (that is the inverse of the constant function $1$ under Dirichlet involution). This allows one to show in particular that
\begin{subequations}
 \begin{equation}
   \sum_{\substack{a = 1 \\ \gcd(a,p)=1}}^p a = \frac{1}{2} p \varphi(p)
 \end{equation}
 \begin{equation}
   \sum_{\substack{a = 1 \\ \gcd(a,p)=1}}^p a^2 = \varphi(p) \left[\frac{1}{3} p^2 + \frac{1}{6} \prod_{\substack{q | p \\ q\, \mathrm{prime}}} (-q)\right]
 \end{equation}
\end{subequations}
for all $p>1$, where the product in the last equation runs over prime factors of $p$. The repartition of the eigenvalues of $\gamma$ given in~\eqref{eq:automorphism_eigenvalues} then leads to the simplification~\eqref{eq:modular_invariance_c} as claimed in section~\ref{subsec:hetdual_modinv}.

\bibliography{bibdual}

\providecommand{\href}[2]{#2}\begingroup\raggedright\begin{thebibliography}{10}

\bibitem{Sen:1995ff}
A.~Sen and C.~Vafa, {\it {Dual pairs of type II string compactification}},
  {\em Nucl. Phys.} {\bf B455} (1995) 165--187,
  [\href{http://xxx.lanl.gov/abs/hep-th/9508064}{{\tt hep-th/9508064}}].

\bibitem{Israel:2013wwa}
{D. Isra\"el} and {V. Thi\'ery}, {\it {Asymmetric Gepner models in type II}},
  {\em JHEP} {\bf 02} (2014) 011,
  [\href{http://xxx.lanl.gov/abs/1310.4116}{{\tt arXiv:1310.4116}}].

\bibitem{Israel:2015efa}
{D. Isra\"el}, {\it {Nongeometric Calabi-Yau compactifications and fractional
  mirror symmetry}},  {\em Phys. Rev.} {\bf D91} (2015) 066005,
  [\href{http://xxx.lanl.gov/abs/1503.0155}{{\tt arXiv:1503.0155}}]. [Erratum:
  Phys. Rev.D91,no.12,129902(2015)].

\bibitem{Blumenhagen:2016axv}
R.~Blumenhagen, M.~Fuchs, and E.~Plauschinn, {\it {Partial SUSY Breaking for
  Asymmetric Gepner Models and Non-geometric Flux Vacua}},  {\em JHEP} {\bf 01}
  (2017) 105, [\href{http://xxx.lanl.gov/abs/1608.0059}{{\tt
  arXiv:1608.0059}}].

\bibitem{Intriligator:1990ua}
K.~A. Intriligator and C.~Vafa, {\it {Landau-Ginzburg orbifolds}},  {\em Nucl.
  Phys.} {\bf B339} (1990) 95--120.

\bibitem{Schellekens:1989wx}
A.~N. Schellekens and S.~Yankielowicz, {\it {New Modular Invariants for $N=2$
  Tensor Products and Four-dimensional Strings}},  {\em Nucl. Phys.} {\bf B330}
  (1990) 103--123.

\bibitem{Hull:2017llx}
C.~Hull, D.~Israel, and A.~Sarti, {\it {Non-geometric Calabi-Yau Backgrounds
  and K3 automorphisms}},  {\em JHEP} {\bf 11} (2017) 084,
  [\href{http://xxx.lanl.gov/abs/1710.0085}{{\tt arXiv:1710.0085}}].

\bibitem{Hull:2004in}
C.~M. Hull, {\it {A Geometry for non-geometric string backgrounds}},  {\em
  JHEP} {\bf 10} (2005) 065,
  [\href{http://xxx.lanl.gov/abs/hep-th/0406102}{{\tt hep-th/0406102}}].

\bibitem{mirror}
P.~Comparin and N.~Priddis, {\it {BHK mirror symmetry for K3 surfaces with
  non-symplectic automorphism}},  \href{http://xxx.lanl.gov/abs/1704.0035}{{\tt
  arXiv:1704.0035}}.

\bibitem{mirror2}
C.~Bott, P.~Comparin, and N.~Priddis, {\it {Mirror symmetry for K3 surfaces}},
  \href{http://xxx.lanl.gov/abs/1901.0937}{{\tt arXiv:1901.0937}}.

\bibitem{Hull:1994ys}
C.~M. Hull and P.~K. Townsend, {\it {Unity of superstring dualities}},  {\em
  Nucl. Phys.} {\bf B438} (1995) 109--137,
  [\href{http://xxx.lanl.gov/abs/hep-th/9410167}{{\tt hep-th/9410167}}].

\bibitem{Ferrara:1995yx}
S.~Ferrara, J.~A. Harvey, A.~Strominger, and C.~Vafa, {\it {Second quantized
  mirror symmetry}},  {\em Phys. Lett.} {\bf B361} (1995) 59--65,
  [\href{http://xxx.lanl.gov/abs/hep-th/9505162}{{\tt hep-th/9505162}}].

\bibitem{Aspinwall:1996mn}
P.~S. Aspinwall, {\it {K3 surfaces and string duality}},  in {\em {Differential
  geometry inspired by string theory}}, pp.~421--540, 1996.
\newblock \href{http://xxx.lanl.gov/abs/hep-th/9611137}{{\tt hep-th/9611137}}.
\newblock [,1(1996)].

\bibitem{Dabholkar:2002sy}
A.~Dabholkar and C.~Hull, {\it {Duality twists, orbifolds, and fluxes}},  {\em
  JHEP} {\bf 09} (2003) 054,
  [\href{http://xxx.lanl.gov/abs/hep-th/0210209}{{\tt hep-th/0210209}}].

\bibitem{Dabholkar:2005ve}
A.~Dabholkar and C.~Hull, {\it {Generalised T-duality and non-geometric
  backgrounds}},  {\em JHEP} {\bf 05} (2006) 009,
  [\href{http://xxx.lanl.gov/abs/hep-th/0512005}{{\tt hep-th/0512005}}].

\bibitem{Harvey:2017xdt}
J.~A. Harvey and G.~W. Moore, {\it {Conway Subgroup Symmetric Compactifications
  of Heterotic String}},  {\em J. Phys.} {\bf A51} (2018), no.~35 354001,
  [\href{http://xxx.lanl.gov/abs/1712.0798}{{\tt arXiv:1712.0798}}].

\bibitem{Vafa:1995gm}
C.~Vafa and E.~Witten, {\it {Dual string pairs with N=1 and N=2 supersymmetry
  in four-dimensions}},  {\em Nucl. Phys. Proc. Suppl.} {\bf 46} (1996)
  225--247, [\href{http://xxx.lanl.gov/abs/hep-th/9507050}{{\tt
  hep-th/9507050}}]. [,225(1995)].

\bibitem{ghi}
Y.~Gautier, C.~Hull, and D.~Israel, {\it {Moduli spaces of non-geometric type
  II/heterotic dual pairs}},  {\em to appear}.

\bibitem{Seiberg:1988pf}
N.~Seiberg, {\it {Observations on the Moduli Space of Superconformal Field
  Theories}},  {\em Nucl. Phys.} {\bf B303} (1988) 286--304.

\bibitem{Aspinwall:1994rg}
P.~S. Aspinwall and D.~R. Morrison, {\it {String theory on K3 surfaces}},
  \href{http://xxx.lanl.gov/abs/hep-th/9404151}{{\tt hep-th/9404151}}.

\bibitem{Nahm:2001kh}
W.~Nahm and K.~Wendland, {\it {Mirror symmetry on Kummer type K3 surfaces}},
  {\em Commun. Math. Phys.} {\bf 243} (2003) 557--582,
  [\href{http://xxx.lanl.gov/abs/hep-th/0106104}{{\tt hep-th/0106104}}].

\bibitem{ReidEdwards:2008rd}
R.~A. Reid-Edwards and B.~Spanjaard, {\it {N=4 Gauged Supergravity from
  Duality-Twist Compactifications of String Theory}},  {\em JHEP} {\bf 12}
  (2008) 052, [\href{http://xxx.lanl.gov/abs/0810.4699}{{\tt
  arXiv:0810.4699}}].

\bibitem{Greene:1990ud}
B.~R. Greene and M.~R. Plesser, {\it {Duality in {Calabi-Yau} Moduli Space}},
  {\em Nucl. Phys.} {\bf B338} (1990) 15--37.

\bibitem{Artebani2014758}
M.~Artebani, {S. Boissi\`ere}, and A.~Sarti, {\it {The
  Berglund-H\"ubsch-Chiodo-Ruan mirror symmetry for K3 surfaces}},  {\em
  Journal de Mathematiques Pures et Appliquees} {\bf 102} (2014), no.~4 758 --
  781.

\bibitem{comparin2014}
P.~Comparin, C.~Lyons, N.~Priddis, and R.~Suggs, {\it {The mirror symmetry of
  K3 surfaces with non-symplectic automorphisms of prime order}},  {\em Adv.
  Theor. Math. Phys.} {\bf 18} (12, 2014) 1335--1368.

\bibitem{Gepner:1987qi}
D.~Gepner, {\it {Space-Time Supersymmetry in Compactified String Theory and
  Superconformal Models}},  {\em Nucl. Phys.} {\bf B296} (1988) 757.
  [,757(1987)].

\bibitem{Witten:1995ex}
E.~Witten, {\it {String theory dynamics in various dimensions}},  {\em Nucl.
  Phys.} {\bf B443} (1995) 85--126,
  [\href{http://xxx.lanl.gov/abs/hep-th/9503124}{{\tt hep-th/9503124}}].
  [,333(1995)].

\bibitem{Aspinwall:1995vk}
P.~S. Aspinwall and J.~Louis, {\it {On the ubiquity of K3 fibrations in string
  duality}},  {\em Phys. Lett.} {\bf B369} (1996) 233--242,
  [\href{http://xxx.lanl.gov/abs/hep-th/9510234}{{\tt hep-th/9510234}}].

\bibitem{Narain:1985jj}
K.~S. Narain, {\it {New Heterotic String Theories in Uncompactified Dimensions
  < 10}},  {\em Phys. Lett.} {\bf 169B} (1986) 41--46.

\bibitem{Narain:1986am}
K.~S. Narain, M.~H. Sarmadi, and E.~Witten, {\it {A Note on Toroidal
  Compactification of Heterotic String Theory}},  {\em Nucl. Phys.} {\bf B279}
  (1987) 369--379.

\bibitem{Harvey:1995rn}
J.~A. Harvey and A.~Strominger, {\it {The heterotic string is a soliton}},
  {\em Nucl. Phys.} {\bf B449} (1995) 535--552,
  [\href{http://xxx.lanl.gov/abs/hep-th/9504047}{{\tt hep-th/9504047}}].
  [Erratum: Nucl. Phys.B458,456(1996)].

\bibitem{GrootNibbelink:2017usl}
S.~Groot~Nibbelink and P.~K.~S. Vaudrevange, {\it {T-duality orbifolds of
  heterotic Narain compactifications}},  {\em JHEP} {\bf 04} (2017) 030,
  [\href{http://xxx.lanl.gov/abs/1703.0532}{{\tt arXiv:1703.0532}}].

\bibitem{Walton:1987bu}
M.~A. Walton, {\it {The Heterotic String on the Simplest Calabi-yau Manifold
  and Its Orbifold Limits}},  {\em Phys. Rev.} {\bf D37} (1988) 377.

\bibitem{Gaberdiel:2012um}
M.~R. Gaberdiel and R.~Volpato, {\it {Mathieu Moonshine and Orbifold K3s}},
  {\em Contrib. Math. Comput. Sci.} {\bf 8} (2014) 109--141,
  [\href{http://xxx.lanl.gov/abs/1206.5143}{{\tt arXiv:1206.5143}}].

\bibitem{vaidyanathaswamy1928integer}
R.~Vaidyanathaswamy, {\it Integer-roots of the unit matrix},  {\em Journal of
  the London Mathematical Society} {\bf 1} (1928), no.~2 121--124.

\bibitem{Narain:1986qm}
K.~S. Narain, M.~H. Sarmadi, and C.~Vafa, {\it {Asymmetric Orbifolds}},  {\em
  Nucl. Phys.} {\bf B288} (1987) 551.

\bibitem{Narain:1990mw}
K.~S. Narain, M.~H. Sarmadi, and C.~Vafa, {\it {Asymmetric orbifolds: Path
  integral and operator formulations}},  {\em Nucl. Phys.} {\bf B356} (1991)
  163--207.

\bibitem{Dabholkar:2005dt}
A.~Dabholkar, F.~Denef, G.~W. Moore, and B.~Pioline, {\it {Precision counting
  of small black holes}},  {\em JHEP} {\bf 10} (2005) 096,
  [\href{http://xxx.lanl.gov/abs/hep-th/0507014}{{\tt hep-th/0507014}}].

\bibitem{Hull:2019iuy}
C.~Hull and R.~J. Szabo, {\it {Noncommutative gauge theories on D-branes in
  non-geometric backgrounds}},  \href{http://xxx.lanl.gov/abs/1903.0494}{{\tt
  arXiv:1903.0494}}.

\bibitem{Kiritsis:1997hj}
E.~Kiritsis, {\em {Introduction to superstring theory}}, vol.~B9 of {\em Leuven
  notes in mathematical and theoretical physics}.
\newblock Leuven U. Press, Leuven, 1998.

\bibitem{Dabholkar:1989jt}
A.~Dabholkar and J.~A. Harvey, {\it {Nonrenormalization of the Superstring
  Tension}},  {\em Phys. Rev. Lett.} {\bf 63} (1989) 478.

\bibitem{Aspinwall:1995mh}
P.~S. Aspinwall, {\it {An N=2 dual pair and a phase transition}},  {\em Nucl.
  Phys.} {\bf B460} (1996) 57--76,
  [\href{http://xxx.lanl.gov/abs/hep-th/9510142}{{\tt hep-th/9510142}}].

\bibitem{Blumenhagen:2013fgp}
R.~Blumenhagen, D.~L{\"u}st, and S.~Theisen, {\em {Basic concepts of string
  theory}}.
\newblock Theoretical and Mathematical Physics. Springer, Heidelberg, Germany,
  2013.

\bibitem{apostol_introduction_1976}
T.~M. Apostol, {\em Introduction to {Analytic} {Number} {Theory}}.
\newblock Undergraduate {Texts} in {Mathematics}. Springer-Verlag, New York,
  1976.

\end{thebibliography}\endgroup

\end{document}